\documentclass[11pt,a4paper]{article}
\usepackage{etex}
\usepackage[utf8]{inputenc}
\usepackage[noblocks]{authblk}

\usepackage{amsmath}
\usepackage{amsfonts}
\usepackage{amssymb}
\usepackage{parskip}
\usepackage{amsthm}
\usepackage{thmtools}
\usepackage{etoolbox}
\usepackage{cite}
\usepackage{color}

\begingroup
    \makeatletter
    \@for\theoremstyle:=definition,remark,plain\do{%
        \expandafter\g@addto@macro\csname th@\theoremstyle\endcsname{%
            \addtolength\thm@preskip\parskip
            }%
        }
\endgroup

\newcommand{\monthname}{%
  \ifcase\the\month
  \or January
  \or February
  \or March
  \or April
  \or May
  \or June
  \or July
  \or August
  \or September
  \or October
  \or November
  \or December
  \fi}

\usepackage{mathtools}

\usepackage[left=3cm,right=3cm,top=3cm,bottom=3cm]{geometry}
\usepackage{array}
\usepackage[all,cmtip]{xy}
\usepackage{graphicx}
\usepackage{tikz,pgfplots}
\usepackage{float}
\usepackage{mathtools}
\usepackage{subcaption}%

\usepackage{booktabs}
\usepackage{ifdraft}

\usepackage{algorithm}
\usepackage{algorithmicx}
\usepackage{algpseudocode}

\usepackage{microtype}	
\usepackage[colorlinks,citecolor=black]{hyperref}
\usepackage{cleveref}

\numberwithin{equation}{section}
\numberwithin{table}{section}
\numberwithin{figure}{section}

\newtheorem{theorem}{Theorem}[section]
\crefname{theorem}{theorem}{theorems}
\Crefname{theorem}{Theorem}{Theorems}

\newtheorem{prop}[theorem]{Proposition}
\crefname{prop}{proposition}{propositions}
\Crefname{prop}{Proposition}{Propositions}

\newtheorem{lemma}[theorem]{Lemma}
\crefname{lemma}{lemma}{lemmas}
\Crefname{lemma}{Lemma}{Lemmas}

\newtheorem{corollary}[theorem]{Corollary}
\crefname{corollary}{corollary}{corollaries}
\Crefname{corollary}{Corollary}{Corollaries}

\theoremstyle{definition}
\newtheorem{remark}[theorem]{Remark}
\AtEndEnvironment{remark}{\hfill\ensuremath{\diamondsuit}}
\crefname{remark}{remark}{remarks}
\Crefname{remark}{Remark}{Remarks}

\theoremstyle{definition}

\crefname{defn}{definition}{definitions}
\Crefname{defn}{Definition}{Definitions}

\newtheorem{example}[theorem]{Example}
\AtEndEnvironment{example}{\hfill\ensuremath{\bigcirc}}
\crefname{example}{example}{example}
\Crefname{example}{Example}{Examples}

\title{Data-driven Structured Realization}
\author{P.~Schulze\thanks{The author was supported by the DFG  Collaborative Research Center 1029 {\it Substantial efficiency increase in gas turbines through direct use of coupled unsteady combustion and flow dynamics}, project A02.}}
\author{B.~Unger\thanks{The author was supported by the DFG Collaborative Research Center 910 {\it Control of self-organizing nonlinear systems: Theoretical methods and concepts of application}, project A2.}}
\affil{Institut f\"ur Mathematik, TU Berlin, Germany, \texttt{$\{$pschulze,unger$\}$@math.tu-berlin.de}.}

\author{C.~Beattie\thanks{The work of this author was supported in part by the Einstein Foundation Berlin.}}
\author{S.~Gugercin\thanks{The work of this author was supported in part by the Alexander von Humboldt Foundation.}}
\affil{Department of Mathematics, Virginia Tech, Blacksburg, VA,  \texttt{$\{$beattie,gugercin$\}$@vt.edu}.}
\newcommand{\preprintNr}{23}

\makeatletter
\let\thetitle\@title
\let\theauthor\@author
\makeatother

\ifdraft{\title{\textsc{\large Working draft}\\\thetitle}}{}

\newcommand{\bfu}{{\boldsymbol{u}}}
\newcommand{\bfb}{{\boldsymbol{b}}}
\newcommand{\bfc}{{\boldsymbol{c}}}
\newcommand{\bfp}{{\boldsymbol{p}}}
\newcommand{\bfx}{{\boldsymbol{x}}}
\newcommand{\bfy}{{\boldsymbol{y}}}

\newcommand{\bff}{{\boldsymbol{f}}}
\newcommand{\bfe}{{\boldsymbol{e}}}
\newcommand{\bfg}{{\boldsymbol{g}}}
\newcommand{\bfv}{{\boldsymbol{v}}}
\newcommand{\bfw}{{\boldsymbol{w}}}

\newcommand{\bfalpha}{{\boldsymbol{\alpha}}}
\newcommand{\bfbeta}{{\boldsymbol{\beta}}}

\newcommand{\original}{Original}
\newcommand{\loewner}{Loewner}
\newcommand{\additional}{Additional points}
\newcommand{\hermite}{Hermite}

\DeclareMathOperator{\diag}{diag}
\DeclareMathOperator{\spann}{span}
\def\numData{n}
\def\leftDir{{\boldsymbol{\ell}}}
\def\LeftDir{\mathcal{L}}
\def\leftPoint{\mu}
\def\LeftPoint{\mathcal{M}}
\def\leftData{{\boldsymbol{f}}}
\def\LeftData{\mathcal{F}}
\def\rightDir{{\boldsymbol{\it r}}}
\def\RightDir{\mathcal{R}}
\def\rightPoint{\sigma}
\def\RightPoint{\mathcal{S}}
\def\rightData{{\boldsymbol{g}}}
\def\RightData{\mathcal{G}}
\def\bitangentialData{\theta}
\def\numFunctions{K}
\def\Hred{\widetilde{H}}

\def\Ared {\widetilde{A}}
\def\Bred{\widetilde{B}}
\def\Cred{\widetilde{C}}
\def\Kred{\widetilde{\mathcal{K}}}
\def\xred{\widetilde{\boldsymbol{x}}}
\def\yred{\widetilde{\boldsymbol{y}}}
\def\dimFOM{N}

\newcommand{\e}[1]{\mathrm{e}^{#1}}

\begin{document}

\ifdraft{}{
	\thispagestyle{empty}
	\begin{center}
		\renewcommand{\thefootnote}{\fnsymbol{footnote}}
		\begin{minipage}{0.29\linewidth}
			\includegraphics[width=1in]{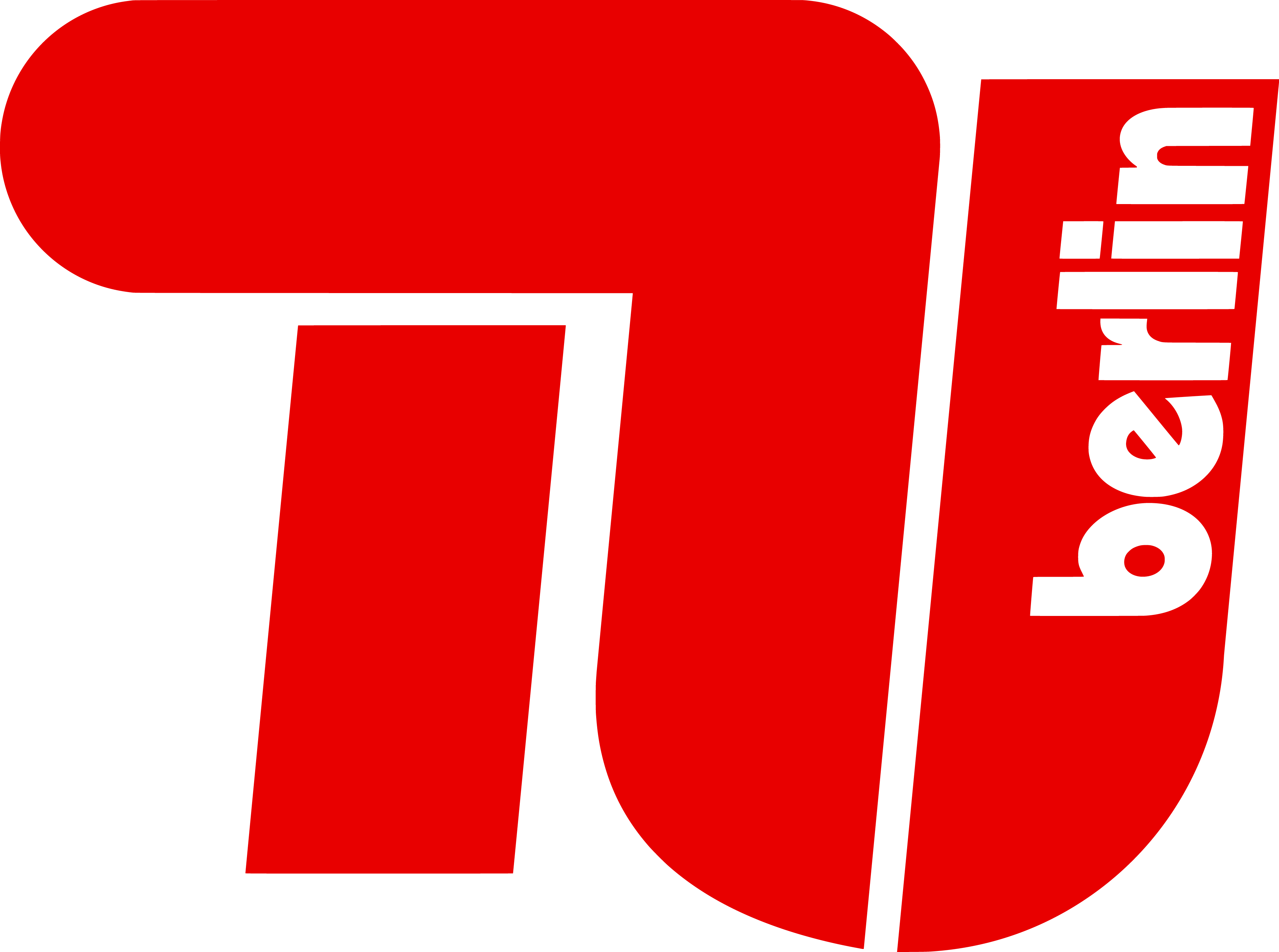}
		\end{minipage}
		\begin{minipage}{0.69\linewidth}
			\begin{flushright}
				{\Huge Technische Universit\"at Berlin}
				\Large Institut f\"ur Mathematik
			\end{flushright}
		\end{minipage}
		\\ [50mm]
		\scalebox{1}{\bf\LARGE \thetitle}\\[13mm]
		{\large\bf 
			\begin{tabular}{ccc}
				Philipp Schulze & \qquad & Benjamin Unger\\ 
				Christopher Beattie &\qquad& Serkan Gugercin
			\end{tabular}
		}\\[18mm]
		{\bf Preprint \preprintNr-\the\year} \\
		\vfill {\bf Preprint-Reihe des Instituts f\"ur Mathematik} \\[1mm]
		{\bf Technische Universit\"at Berlin} \\[1mm]
		{\bf \tt http://www.math.tu-berlin.de/preprints} \\[8mm]
		{\bf Preprint \preprintNr-\the\year \hfill \monthname\ \the\year}
	\end{center}
	\newpage
	\setcounter{page}{1}
}

\maketitle

\begin{abstract} 
We present a framework for constructing structured realizations of linear dynamical systems having transfer functions of the form $\Cred(\sum_{k=1}^K h_k(s)\Ared_k)^{-1}\Bred$ where $h_1,\,h_2,\,...,h_K$ are prescribed functions that specify the surmised structure of the model. Our construction is data-driven in the sense that an interpolant is derived entirely from measurements of a transfer function. Our approach extends the Loewner realization framework to more general system structure that includes  second-order (and higher) systems as well as systems with internal delays. Numerical examples demonstrate the advantages of this approach.
\end{abstract}
\vskip .3truecm
\noindent
{\bf AMS(MOS) subject classification:} 93B15, 30E05, 93C05
\section{Introduction}
The simulation of complex physical, chemical, or biological processes is a standard task in science, engineering, and industry. 
The dynamics of such processes are commonly modeled as dynamical systems, which then can be analyzed (often through simulation) for optimization and control. The demand for higher fidelity models produces as a common consequence  ever more complex and larger dynamical systems, whose simulation may require computational resources that become unmanageably large.  This computational cost is often directly related to the state space dimension of the underlying dynamical system, thus creating a need for low-dimensional approximations of large-scale models. Model order reduction (MOR) techniques using rational interpolation methods, such as the \emph{iterative rational Krylov algorithm} (IRKA) \cite{GuGAB08}, or Gramian-based methods, such as balanced truncation
\cite{MulR76,Moo81},
have become popular tools to create such reduced-order models (ROMs); for an overview of these approaches, see the survey papers \cite{AntSG01,AntBG10,BauBF14} and the monograph \cite{Ant05}. 
There are many features that distinguish model reduction methods from one another; we focus on the dichotomy between \emph{projection-based methods} and \emph{data-driven methods}.  
Projection-based methods presuppose access to an explicit state space realization of the original dynamical system and then
identify low-dimensional, high-value subspaces of the state space, using projections to constrain dynamics to these subspaces. 
Data-driven model reduction methods 
as exemplified by \emph{vector fitting} \cite{GusS99,DrmGB14} or the \emph{Loewner realization} framework of \cite{MayA07} are  \emph{nonintrusive} in the sense that the access to internal dynamics that an explicit state space realization provides is not necessary. Such methods are able to produce system realizations (generally low-dimensional) directly from measurements of the transfer function. The greater flexibility that data-driven methods provide is balanced against 
the general inability of data-driven methods to preserve structural features that may be present in the original model, a capability that often is available to projection-based methods.   

In some practical settings, the original (or approximate) model may be available only implicitly either through response observation or simulation, leading one to data-driven approaches as the only feasible strategy for MOR.   Despite the inaccessibility of a description of detailed internal dynamics,  there may yet be significant ancillary information or at least a good basic understanding of how the system should behave, allowing one to surmise general structural features of the underlying dynamical system.  For example, vibration effects are naturally associated with subsystems that have second-order structure;  internal transport or signal propagation will naturally be associated with state delays.  

\begin{example}
	\label{acousticWaveExample}
	Consider acoustic transmission in a fluid-filled duct of length $L$ that has an acoustic driver positioned at one end. Suppose we are interested in the acoustic pressure $y(t)=p(\xi_0,t)$ at a fixed point $\xi_0\in(0,L)$ in the duct, which we view as the output of an abstract system that is driven by the input fluid velocity $u(t)$, determined by an acoustic driver positioned at $\xi=0$. We assume that the output pressure depends linearly on the input velocity in a way that is invariant to translation in time, and so the output could be anticipated to involve some superposition of internal states that are lagged in time according to propagation delays related to the distance traveled by the signal.  Assuming a uniform sound speed $c$ throughout the duct, we allow for a direct propagation delay $\tau_1 = \xi_0/c$ between the input and output location and a second propagation delay $\tau_2=(2L-\xi_0)/c$, associated with a reflected signal. A semi-empirical model for the state evolution of a system that has these basic features could have the form
	\begin{displaymath}
 		A_1\bfx(t) + A_2\bfx(t-\tau_1) +A_3\bfx(t-\tau_2) = \bfb u(t),
	\end{displaymath}
with an output port map given by $y(t)=\bfc^T\,\bfx(t)$. The matrices $A_1$,  $A_2$, and  $A_3$, the port maps associated with the vectors, $\bfb$ and $\bfc$, as well as their dimensions are unknown. We arrive at a (semi-empirical) transfer function for this system having the hypothesized structure
	\begin{displaymath}
		H(s)=\bfc^T\left( A_1 + A_2 \mathrm{e}^{-\tau_1s} +A_3\, \mathrm{e}^{-\tau_2s}\right)^{-1}\bfb.
	\end{displaymath} 
Based on observed or computed system response data, we wish to identify $A_1$,  $A_2$,  $A_3$, $\bfb$ and $\bfc$. 
\end{example}

We present here a general data-driven model reduction framework that is capable of preserving system structure present in an original  model when it is known, and possibly inducing hypothesized structure in other cases.   We lay out our basic problem framework in the following Section \ref{sec:probSetting} and then show how to exploit this in developing structured realizations in Section \ref{sec:StructureLoewner}.  Computational examples are offered in Section \ref{sec:examples}.

\section{Problem Setting} \label{sec:probSetting}

Although the term \emph{system structure} can have wide ranging meanings, for our purposes we will understand 
the term to refer to equivalence classes of systems having realizations associated with a linearly independent function family $\{h_1,\, h_2,\, \ldots,\, h_\numFunctions\}$ that appear as 
\begin{equation}
	\label{eq:FOMstructured}
	H(s) = C\left(\sum_{k=1}^\numFunctions h_k(s) A_k\right)^{-1}B,
\end{equation}
where 
$C\in\mathbb{R}^{p\times\dimFOM}$, $A_k\in\mathbb{R}^{\dimFOM\times\dimFOM}$ for $k=1,\ldots,\numFunctions$, $B\in\mathbb{R}^{\dimFOM\times m}$. We assume in all that follows that the functions involved, $h_k:\mathbb{C}\to\mathbb{C}$, are meromorphic. 
For any given function family, we will refer to associated matrix-valued functions having the form $\sum_{k=1}^\numFunctions h_k(s) A_k$ 
as an \emph{affine structure}.  By standard abuse of notation, we use $H(s)$ to denote either the system itself or the transfer function of the system evaluated at the point $s\in\mathbb{C}$. The two systems $H(s)$ and $\Hred(s)$ are called \emph{structurally equivalent} if $H(s),\,\Hred(s)\in\mathbb{C}^{p\times m}$ for $s\in\mathbb{C}$ and if they each have the form
\begin{displaymath}
	H(s) = C\left(\sum_{k=1}^\numFunctions h_k(s) A_k\right)^{-1}B
 	\quad\mbox{ and } \quad
 	\Hred(s) = \Cred\left(\sum_{k=1}^\numFunctions \tilde{h}_k(s) \Ared_k\right)^{-1}\Bred,
\end{displaymath}
 with
$\spann \{h_1,\, h_2,\, \ldots,\, h_\numFunctions\} \equiv \spann \{\tilde{h}_1,\, \tilde{h}_2,\, \ldots,\, \tilde{h}_{\numFunctions}\}$.  
In particular, we allow different state space dimensions, i.\,e., for  $\Cred\in\mathbb{R}^{p\times\numData}$, $\Ared_k\in\mathbb{R}^{\numData\times\numData}$, and $\Bred\in\mathbb{R}^{\numData\times m}$
the integers $\dimFOM$ and $\numData$ need not be the same.  
Given an original (full order) system associated with $H(s)$, we aim to construct a structurally equivalent system $\Hred(s)$, with state space dimension $\numData \ll \dimFOM$, and we wish to accomplish this allowing only evaluations of $H(s)$. 

The general structure of \eqref{eq:FOMstructured} encompasses a variety of system formulations.
Indeed, we may observe immediately that linear time invariant dynamical systems often are presented as standard first-order realizations
\begin{equation}
	\label{eq:stateSpaceSystem}
		A_1\dot{\bfx}(t) + A_2\bfx(t) = B\bfu(t),\qquad 
	  	\bfy(t) = C\bfx(t),
\end{equation}
where $A_1, A_2\in\mathbb{R}^{\dimFOM\times\dimFOM}$, $B\in\mathbb{R}^{\dimFOM\times m}$, and $C\in\mathbb{R}^{p\times\dimFOM}$. The \emph{state} is denoted by $\bfx(t)\in\mathbb{R}^\dimFOM$, while $\bfu(t)\in\mathbb{R}^m$ and $\bfy(t)\in \mathbb{R}^p$ are, respectively, the \emph{inputs} and \emph{outputs} of the system. 
 If the input $u:[0,T]\to\mathbb{R}^m$  is exponentially bounded, i.\,e., $\|u(t)\|\in\mathcal{O}(e^{\gamma t})$ 
 for some $\gamma\in\mathbb{R}$,  then $x:[0,T]\to\mathbb{R}^n$ and 
 $y:[0,T]\to\mathbb{R}^p$ are also exponentially bounded.
  The Laplace transform may be applied to \eqref{eq:stateSpaceSystem} and rearranged to 
$	\hat{\bfy}(s) = H(s)\hat{\bfu}(s) $
with $H(s) = C(sA_1-A_2)^{-1}B$, which has the form of  \eqref{eq:FOMstructured} with $h_1(s)=s$ and $h_2(s)\equiv-1$.  

In many practical applications, the underlying dynamical system comes in a form quite different from \eqref{eq:stateSpaceSystem} that will reflect the nature of the system, and we may wish to preserve this structure in the realization. For example, a general RLC network may be modeled as integro differential-algebraic equation \cite{Fre08}, given by
\begin{equation}
	\label{eq:RLC}
		A_1\dot{\bfx}(t) + A_2 \bfx(t) + A_3 \int_0^t \bfx(\tau) d\tau = B\bfu(t),\qquad \bfy(t) = B^T \bfx(t).
\end{equation} 
The transfer function associated with \eqref{eq:RLC} is given by
\begin{equation*}
	H(s) = B^T\left(sA_1 + A_2 + \frac{1}{s}A_3\right)^{-1}B
\end{equation*}
and we expect better approximation properties of the ROM by preserving this form. Further examples for system structures are listed in \Cref{tab:StructureExamples}.
\begin{table}
	\centering
	\caption{Different system structures (all with output mapping $\bfy(t) = C\bfx(t)$)}
	\label{tab:StructureExamples}
	{\footnotesize
	\begin{tabular}{lcc}
		& \textbf{state space description} & \textbf{transfer function}\\\toprule
		second-order & $A_1\ddot{\bfx}(t) + A_2\dot{\bfx}(t) + A_3\bfx(t) = B\bfu(t)$ & $C\left(s^2A_1 + sA_2 + A_3\right)^{-1}B$\\\midrule
		state delay & $A_1\dot{\bfx}(t) + A_2\bfx(t) + A_3\bfx(t-\tau) = B\bfu(t)$ & $C\left(sA_1 + A_2 + \mathrm{e}^{-\tau s}A_3\right)^{-1}B$\\\midrule
		neutral delay & $A_1\dot{\bfx}(t) + A_2\bfx + A_3\dot{\bfx}(t-\tau) = B\bfu(t)$ & $C\left(sA_1 + A_2 + s\mathrm{e}^{-\tau s}A_3\right)^{-1}B$\\\midrule
		viscoelastic & $A_1\ddot{\bfx}(t) + \int_0^t \mathsf{h}(t\!-\!\tau)A_2\dot{\bfx}(\tau)\mathrm{d}\tau + A_3\bfx(t) = B\bfu(t)$ & $C\left(s^2A_1 + s\hat{h}(s)A_2 + A_3\right)^{-1}\!B$\\\bottomrule
	\end{tabular}
	}
\end{table}

Preserving structure often allows one to derive reduced models with smaller \emph{state space dimension} $\numData$, 
while maintaining comparable or at times even better accuracy than what unstructured reduced models produce, see Section~5 in \cite{BeaG09}.  Additionally, since the internal structure of models often reflects core phenomenological properties, structured models may behave in ways that remain qualitatively consistent with the phenomena that are being modeled -- possibly more so than unstructured models having higher objective fidelity.

Structure-preserving model reduction has for the most part developed in a projection-based context that presupposes access to internal dynamics.  Projection-based techniques are often able to retain special structural features in the reduced models that may reflect underlying physical properties of the systems under study \cite{SuC91,MeyS96,BeaG09,ChaBG16,Fre08,LalKM03,ChaGVV05}.   
Data-driven techniques for system identification and model reduction do not generally have this capacity, however, the recent contributions of \cite{SchU15,PonPS15} provide a notable exception for time-delay systems.  In the present work, we build on the results of \cite{SchU15} and extend its domain of applicability to a wide range of structured dynamical systems.  These ideas originate in the Loewner realization framework of \cite{MayA07}.

The Loewner realization framework is an effective and broadly applicable approach for constructing rational approximants directly from interpolation data; it has been extended to parametric systems \cite{IonA14,AntIL12} and to realization independent methods for optimal $\mathcal{H}_2$ approximation \cite{BeaG12}. However, the Loewner framework is only capable of producing \emph{rational} approximants and, so in particular, it cannot capture the transcendental character of transfer functions for dynamical systems containing distributed parameter subsystems that model convection or transport (cf.  \cite{CurM09}).   

We begin by describing in detail the type of system response data that we will assume henceforth to be available. 
Suppose we have $2\numData$ points in the complex plane, which may be interpreted as complex driving frequencies, 
$\{\leftPoint_1,\ldots,\leftPoint_\numData\}$ and $\{\rightPoint_1,\ldots,\rightPoint_\numData\}$. We assume for the time being that these two sets, $\{\leftPoint_i\}_{i=1}^{\numData}$ and $\{\rightPoint_i\}_{i=1}^{\numData}$, are each made up of $\numData$ distinct points, although we allow the two point sets to have nontrivial intersection (so that it could happen that $\leftPoint_i=\rightPoint_j$ for some index pairs $(i,j)$). In addition to these complex frequencies, we have the so-called \emph{left tangential direction} vectors $\{\leftDir_1,\ldots,\leftDir_\numData\}$ and 
the \emph{right tangential direction} vectors $\{\rightDir_1,\ldots,\rightDir_\numData\}$ where $\leftDir_i \in \mathbb{R}^p$ and $\rightDir_i \in \mathbb{R}^m$
for $i=1,\ldots,n$. In the single-input/single-output (SISO) case, these tangential directions are assigned the value one, i.\,e., $\leftDir_i = \rightDir_i = 1$. Unlike  projection-based model reduction, which requires access to the state space quantities,  data-driven interpolatory model reduction only assumes access to the action of the transfer function evaluated at the driving frequencies along the tangential directions, i.\,e., 
\begin{align}
\label{eq:dataSamplingOfTransferFunction}
	\leftDir_i^T H(\leftPoint_i) &= \leftData_i^T &\text{and}&&
	H(\rightPoint_i)\rightDir_i &= \rightData_i \qquad\text{for } i=1,\ldots,\numData.
\end{align}
If the direction vectors $\leftDir_i$ and $\leftDir_j$ are linearly independent, one can allow $\leftPoint_i$ to coincide with $\leftPoint_j$, and similarly for $\rightPoint_i$'s. However, for simplicity the only coincidence of interpolation points that we admit will be between left and right interpolation points, i.\,e., $\leftPoint_i=\rightPoint_j$. If this is the case for an index pair $(i,j)$, then bitangential derivative data is assumed to be available. Since we assume that each of the two sets  $\{\leftPoint_i\}_{i=1}^{\numData}$ and $\{\rightPoint_i\}_{i=1}^{\numData}$ consists of $\numData$ distinct points,  if $\leftPoint_i=\rightPoint_j$ for an index pair $(i,j)$, without loss of generality, we assume $i=j$. Then, the corresponding bitangential derivative data is defined as
$$
\leftDir_i^T H'(\leftPoint_i) \rightDir_i = \bitangentialData_i.
$$
Following \cite{MayA07,AntBG10}, we summarize the \emph{interpolation data} as
\begin{align}
	\mbox{\small left interpolation data:}\ & \{(\leftPoint_i,\leftDir_i,\leftData_i)\ |\ \leftPoint_i\in\mathbb{C}, \leftDir_i\in\mathbb{C}^{p}, \leftData_i\in\mathbb{C}^{m}, i=1,\ldots,\numData\},\nonumber\\	
	\mbox{\small right interpolation data:}\ & \{(\rightPoint_i,\rightDir_i,\rightData_i)\ |\ \rightPoint_i\in\mathbb{C}, \rightDir_i\in\mathbb{C}^m, \rightData_i\in\mathbb{C}^p, i=1,\ldots,\numData\},	 	\label{eq:interpolationData}	 \\
		\mbox{\small bitangential derivative data:}\ & \{(i,\bitangentialData_i)\ |\ i \in\{1,\ldots,\numData\}\ \mbox{\small for which } \leftPoint_i = \rightPoint_i, \bitangentialData_i \in\mathbb{C} \},	\nonumber
\end{align}
with the understanding that the last category may be empty if $\{\leftPoint_i\}_{i=1}^{\numData} \cap \{\rightPoint_i\}_{i=1}^{\numData}= \varnothing$.
Note that in the case $\leftPoint_i=\rightPoint_i$, the compatibility of the conditions \eqref{eq:dataSamplingOfTransferFunction} requires 
that $\leftData_i^T \rightDir_i = \leftDir_i^T\rightData_i$. 

For ease of presentation, we introduce the matrices 
\begin{align*}
	\LeftPoint &\vcentcolon= \text{diag}(\leftPoint_1,\ldots,\leftPoint_\numData)\in\mathbb{C}^{\numData\times\numData}, & \RightPoint &\vcentcolon= \text{diag}(\rightPoint_1,\ldots,\rightPoint_\numData)\in\mathbb{C}^{\numData\times\numData}, \\
	\LeftDir &\vcentcolon= \begin{bmatrix}
		\leftDir_1 & \ldots & \leftDir_\numData
	\end{bmatrix}\in\mathbb{C}^{p\times\numData}, & \RightDir &\vcentcolon= \begin{bmatrix}
		\rightDir_1 & \ldots & \rightDir_\numData
	\end{bmatrix}\in\mathbb{C}^{m\times\numData},\\
	\LeftData &\vcentcolon= \begin{bmatrix}
		\leftData_1 & \ldots & \leftData_\numData
	\end{bmatrix}\in\mathbb{C}^{m\times\numData}, &
	\RightData &\vcentcolon= \begin{bmatrix}
		\rightData_1 & \ldots & \rightData_\numData
	\end{bmatrix}\in\mathbb{C}^{p\times\numData}.
\end{align*}
Our  goal is to construct matrices $\Cred, \Ared_k$, and $\Bred$ (for $k=1,\ldots,\numFunctions$) using the measurements \eqref{eq:interpolationData}, such that the transfer function $\Hred(s) = \Cred(\sum_{k=1}^\numFunctions h_k(s)\Ared_k)^{-1}\Bred$ satisfies the \emph{interpolation conditions}
\begin{subequations}
\label{eq:allInterpolationConditions}
\begin{equation}
	\label{eq:interpolationCondition}
	\leftDir_i^T\Hred(\leftPoint_i) = \leftDir_i^T H(\leftPoint_i) = \leftData_i^T\quad\text{and}\quad \Hred(\rightPoint_i)\rightDir_i = H(\rightPoint_i)\rightDir_i = \rightData_i\qquad\text{for } i=1,\ldots,\numData.
\end{equation}
If $\leftPoint_i=\rightPoint_i$ for any index $i$, then additionally,
\begin{equation}
	\label{eq:bitangentialInterpCond}
	\leftDir_i^T \Hred     '(\leftPoint_i) \rightDir_i = \leftDir_i^T H'(\leftPoint_i) \rightDir_i = \theta_i
\end{equation}
\end{subequations}
is to be satisfied.

\section{Structured Interpolatory Realizations}
\label{sec:StructureLoewner}
\subsection{Standard Loewner Realization}
A key tool for our results is the Loewner realization framework introduced in \cite{MayA07}. This framework uses the Loewner matrix $\mathbb{L} \in \mathbb{C}^{n \times n}$ 
and the shifted Loewner matrix $\mathbb{L}_\sigma  \in \mathbb{C}^{n \times n}$, whose entries
$\left[\mathbb{L}\right]_{i,j}$ and $\left[\mathbb{L}_{\sigma}\right]_{i,j}$ for $i,j=1,\ldots,\numData$ are defined as
\begin{align}
&\left[\mathbb{L}\right]_{i,j} = \frac{\leftData_i^T\rightDir_j-\leftDir_i^T\rightData_j}{\leftPoint_i-\rightPoint_j}\quad\text{and}\quad
		\left[\mathbb{L}_{\sigma}\right]_{i,j} = \frac{\leftPoint_i\leftData_i^T\rightDir_j-\rightPoint_j\leftDir_i^T\rightData_j}{\leftPoint_i-\rightPoint_j},\quad \mbox{if }\quad \leftPoint_i\neq\rightPoint_j,
			\label{eq:LoewnerMatrix} \\
&\left[\mathbb{L}\right]_{i,i} =\bitangentialData_i \quad\hspace{12.3ex}\text{and}\quad
		\left[\mathbb{L}_{\sigma}\right]_{i,i} = \leftData_i^T\rightDir_i +\leftPoint_i \bitangentialData_i, \hspace{8.1ex}\mbox{if }\quad \leftPoint_i=\rightPoint_i.
			\label{eq:LoewnerMatrixDeriv}
\end{align}

For SISO systems, $\mathbb{L}$ and $\mathbb{L}_{\sigma}$ are the divided differences matrices corresponding to the transfer functions $H(s)$ and $sH(s)$, respectively.
\begin{theorem}[Loewner realization \cite{MayA07}]
	\label{thm:LoewnerRealization}
	Let $\det(\tilde{s}\mathbb{L}-\mathbb{L}_{\sigma})\neq 0$ for all $\tilde{s}\in\{\leftPoint_i\}_{i=1}^n\cup\{\rightPoint_i\}_{i=1}^n$. Then the system
	\begin{equation}
		\label{eq:LoewnerRealization}
		-\mathbb{L}\dot{\xred}(t) = -\mathbb{L}_\sigma \xred(t) + \LeftData^T \bfu(t),\qquad
		\yred(t) = \RightData \xred(t) 
	\end{equation}
	is a minimal realization of an interpolant of the data, i.\,e., its transfer function
	\begin{displaymath}
		\Hred(s) = \RightData(\mathbb{L}_\sigma - s\mathbb{L})^{-1}\LeftData^T
	\end{displaymath}
	satisfies the interpolation conditions \eqref{eq:allInterpolationConditions}.
\end{theorem}
The condition $\det(\tilde{s}\mathbb{L}-\mathbb{L}_{\sigma})\neq 0$ in \Cref{thm:LoewnerRealization} can be relaxed by means of the short singular value decomposition (SVD) \cite[Remark 3.2.1]{Ant05}. If an $\tilde{s}\in\{\leftPoint_i\}_{i=1}^n\cup\{\rightPoint_i\}_{i=1}^n$ violates the regularity condition, then the short SVD of $\tilde{s}\mathbb{L}-\mathbb{L}_\sigma$ can be used to truncate the redundant parts \cite{MayA07}.
%

\subsection{Interpolation Conditions}
\label{subsec:interpConditions}
Suppose we are given interpolation data as in \eqref{eq:interpolationData} and for the moment assume that we already have a realization of the form $\Hred(s) = \Cred\Kred(s)^{-1}\Bred$. If we can impose conditions on $\Cred, \Bred$ and the matrix function $\Kred$ such that $\Hred(s) = \Cred\Kred(s)^{-1}\Bred$ satisfies the interpolation conditions \eqref{eq:allInterpolationConditions}, then we can revert the process and use the conditions to construct the realization. The following observation, which corresponds to an equivalent parametrization of the interpolation conditions \eqref{eq:allInterpolationConditions},  suggests how one might proceed. This  parametrization will form the basis for constructing the structured interpolatory realizations.

\begin{theorem}
	\label{thm:interpolationConditions}
	Let $\Kred(s)$ be a continuously differentiable $\numData\times\numData$ matrix-valued function of the complex argument $s$, which is nonsingular at $s=\leftPoint_i$ and $s=\rightPoint_j$ for $i,j=1,\ldots,\numData$.  
	The realization $\Hred(s) = \Cred\Kred(s)^{-1}\Bred$  satisfies the interpolation conditions \eqref{eq:interpolationCondition}
	if and only if
\begin{equation}  \label{eq:cond1}
	\RightData = \Cred P_{\RightData}~~~\mbox{and}~~~ \LeftData^T = P_{\LeftData}^T\Bred, 
\end{equation}
where $P_{\RightData},P_{\LeftData} \in\mathbb{C}^{\numData\times\numData}$ are two matrices, whose columns $\bfp_{\RightData}^i \vcentcolon= P_{\RightData}\bfe_i$ and
	 $\bfp_{\LeftData}^i \vcentcolon= P_{\LeftData}\bfe_i$, respectively, solve the linear systems
	\begin{equation}  \label{eq:defineP}
		\Kred(\rightPoint_i)\bfp_{\RightData}^i = \Bred\rightDir_i \qquad\text{and}\qquad  \Kred(\leftPoint_i)^T\bfp_{\LeftData}^i = \Cred^T\leftDir_i,
	\end{equation}
	where $\bfe_i$ is the $ith$ column of the $\numData\times\numData$ identity matrix.
	If additionally, $\leftPoint_i = \rightPoint_i$, then $\Hred(s) = \Cred\Kred(s)^{-1}\Bred$ satisfies the bitangential interpolation condition \eqref{eq:bitangentialInterpCond} as well provided that
	\begin{equation}  \label{eq:cond2}
	\left(\bfp_{\LeftData}^i\right)^T\Kred'(\leftPoint_i)\bfp_{\RightData}^i = - \bitangentialData_i.
\end{equation}
\end{theorem}

\begin{proof} 
	The transfer function $\Hred(s) = \Cred\Kred(s)^{-1}\Bred$ is well-defined at $s=\leftPoint_i$ and $s=\rightPoint_i$. 
	Assume \eqref{eq:cond1} and \eqref{eq:defineP}.
Multiplying the first equation in  \eqref{eq:cond1} by $\bfe_i$ yields 
$\bfg_i = \Cred \bfp_{\RightData}^i$. Then, using the first equation in
\eqref{eq:defineP} and the fact that $\Kred(\sigma_i)$ is invertible, one immediately obtains $\bfg_i = \Hred(\sigma_i) \rightDir_i$, i.\,e., the right tangential interpolation holds. Similarly, using the second expression in \eqref{eq:cond1}  and the definition of  $\bfp_{\LeftData}^i$ in \eqref{eq:defineP}, we arrive at $\bff_i^T = \leftDir_i^T\Hred(\mu_i)$; thus \eqref{eq:interpolationCondition} holds. The other direction follows directly. 
	Moreover, if $\leftPoint_i = \rightPoint_i$, then \eqref{eq:cond2} yields
	\begin{displaymath}
		\leftDir_i^T \Hred^{\prime}(\leftPoint_i)\rightDir_i 
= -\leftDir_i^T \Cred\Kred(\leftPoint_i)^{-1}\Kred'(\leftPoint_i)\Kred(\rightPoint_i)^{-1}\Bred\rightDir_i 
= -\left(\bfp_{\LeftData}^i\right)^T \Kred'(\leftPoint_i) \bfp_{\RightData}^i = \bitangentialData_i.
	\end{displaymath}
\end{proof}

Evidently, in order to satisfy the collected tangent interpolation conditions \eqref{eq:interpolationCondition}, we can now equivalently require the realization
$\Hred(s)$ to satisfy the conditions of \Cref{thm:interpolationConditions}. In particular we need $\Kred(s)$ to be nonsingular at the driving frequencies $s=\leftPoint_i$ and $s=\rightPoint_j$. For $\Kred(s) = \sum_{k=1}^\numFunctions h_k(s)\Ared_k$, the other conditions \eqref{eq:cond1}  and \eqref{eq:defineP} can be rewritten as
\begin{align}
	\label{eq:Pequations}
	\RightData &= \Cred P_{\RightData},
	 & \LeftData^T &= P_{\LeftData}^T\Bred,\\
	\label{eq:matrixConditions} \sum_{k=1}^\numFunctions \Ared_k P_{\RightData} h_k(\RightPoint) &= \Bred\RightDir,&    \sum_{k=1}^\numFunctions h_k(\LeftPoint) P_{\LeftData}^T \Ared_k &= \LeftDir^T\Cred,
\end{align}
where we set $h_k(\LeftPoint) \vcentcolon= \diag(h_k(\leftPoint_1),\ldots,h_k(\leftPoint_\numData))$ and $h_k(\RightPoint) \vcentcolon=\diag(h_k(\rightPoint_1),\ldots,h_k(\rightPoint_\numData))$.
To fulfill additionally the bitangential interpolation conditions \eqref{eq:bitangentialInterpCond} for the case $\leftPoint_i=\rightPoint_i$, the third condition of \Cref{thm:interpolationConditions} needs to be satisfied.

If the matrices $P_{\LeftData}$ and $P_{\RightData}$ are nonsingular, then
\begin{displaymath}
	\Hred(s) = \Cred\Kred(s)^{-1}\Bred = \RightData\left(P_{\LeftData}^T\Kred(s)P_{\RightData}\right)^{-1}\LeftData^T,
\end{displaymath}
and hence the realization is unique up to the state space transformation described by $P_{\LeftData}$ and $P_{\RightData}$. In this case, the matrices $\Bred$ and $\Cred$ are given directly by the data without further computations and the matrices $P_{\LeftData}$ and $P_{\RightData}$ capture the non-uniqueness of the realization. In \Cref{subsec:K3} we will use these matrices to tailor the realization to interpolate additional data. In any case, we view equations \eqref{eq:Pequations} and \eqref{eq:matrixConditions} not as a coupled system but as a staggered process. First, fix matrices $P_{\LeftData}, P_{\RightData}$ and determine $\Bred$ and $\Cred$ from \eqref{eq:Pequations}. Then, in the second step, use this information to solve \eqref{eq:matrixConditions}. With this viewpoint, i.\,e., not counting $P_{\LeftData}$ and $P_{\RightData}$ as unknowns, we have $\numFunctions\numData^2$ unknowns from the coefficient matrices $\Ared_k$ and $(m+p)\numData$ unknowns from the input and output matrices $\Bred$ and $\Cred$, giving a total of $\numFunctions\numData^2 + (m+p)\numData$ unknowns. For these unknowns, \eqref{eq:Pequations} and \eqref{eq:matrixConditions} constitute $2\numData^2 + (m+p)\numData$ equations, leaving $(\numFunctions-2)\numData^2$ degrees of freedom. In particular, we can expect a unique solution for $\numFunctions=2$.
\begin{remark}
\label{rem:Kis1} 
\textcolor{black}{We have $(\numFunctions-2)\numData^2$ degrees of freedom to solve the structured realization problem, and therefore
the $\numFunctions = 1$ case does not have enough degrees of freedom to guarantee a solution in general.
Note that the standard rational approximation has $K=2$.  
}
To further examine the $K=1$ case, assume for simplicity that $H(s)$ is SISO, i.\,e., $\Bred = \bfb \in\mathbb{R}^\numData$ and $\Cred^T = \bfc\in\mathbb{R}^\numData$. Then, for $\numFunctions = 1$, the reduced model has the form 
$\Hred(s) = \frac{1}{h_1(s)} \bfc^T \Ared^{-1} \bfb$. Therefore, the interpolation conditions become 
\begin{equation} \label{eq:sisokis1}
\bfc^T \Ared^{-1} \bfb = H(\sigma_i) h_1(\sigma_i)~~~ \mbox{and}~~~  \bfc^T \Ared^{-1} \bfb = H(\mu_i) h_1(\mu_i),~~~
\mbox{for}~~~i=1,\ldots,\numData.
\end{equation}
Since $\bfc^T \Ared^{-1} \bfb$ is constant, for the interpolation problem in \eqref{eq:sisokis1} to have a solution, we need $H(\sigma_i) h_1(\sigma_i) =  H(\mu_i) h_1(\mu_i) = \mathsf{c}$ where $\mathsf{c}$ is a constant for $i=1,\ldots,\numData$. This clearly will not be the case in general and we cannot expect to have a solution. Interestingly, if this condition holds, a solution can be found easily by setting $\Ared = I_n$, $\bfb = \bfe_1$ and $\bfc =   \mathsf{c} \cdot \bfe_1$.
\textcolor{black}{Based on these considerations, we will focus on $K \geq 2$ in the rest of the paper.}
\end{remark}

\begin{remark}
	The nonsingularity of the matrices $P_{\LeftData}$ and $P_{\RightData}$ is connected to the minimality of the realization. To see this, assume that we have a SISO standard state space system, i.\,e., $\Kred(s) = sI_\numData - \Ared$, $\Bred = \bfb\in\mathbb{R}^\numData$, and $\Cred^T = \bfc\in\mathbb{R}^\numData$. Note that in this case $\Kred$ and its pointwise inverse form a set of commutative matrices. Hence we have
	\begin{align*}
		\mathrm{rank}\left(P_{\RightData}\right) &= \mathrm{rank}\left(\begin{bmatrix}
			\Kred(\rightPoint_1)^{-1}\bfb & \cdots & \Kred(\rightPoint_n)^{-1}\bfb
    		\end{bmatrix}\right)\\
		 &= \mathrm{rank}\left(\begin{bmatrix}
		 	\Kred(\rightPoint_1)^{-1}\bfb &  \Kred(\rightPoint_1)^{-1}\Kred(\rightPoint_2)^{-1}\bfb & \cdots & \left(\prod_{i=1}^n \Kred(\rightPoint_i)^{-1}\right)\bfb
		 \end{bmatrix}\right) \\
		 &= \mathrm{rank}\left(\begin{bmatrix}
		 	\bfb & \Ared\bfb & \cdots & \Ared^{\numData-1}\bfb
		 \end{bmatrix}\right)
	\end{align*}
	such that $P_{\RightData}$ is nonsingular if and only if the realization is controllable. Similarly, $P_{\LeftData}$ is nonsingular if and only if the realization is observable.
\end{remark}

%

Note that the Loewner pencil with $h_1(s) \equiv 1$ and $h_2(s) = -s$ satisfies the conditions of \Cref{thm:interpolationConditions} with $\widetilde{\mathcal{K}}(s) =\mathbb{L}_{\sigma}- s\mathbb{L}$, i.\,e., the Loewner framework works with matrices $P_{\LeftData}$ and $P_{\RightData}$ being the identity. Indeed, for $\leftPoint_i\neq\rightPoint_j$, the $(i,j)$ component of the Loewner pencil is
\begin{displaymath}
	\bfe_i^T \widetilde{\mathcal{K}}(s)\bfe_j = \left[\mathbb{L}_{\sigma}\right]_{i,j}- s\left[\mathbb{L}\right]_{i,j} = \left(\frac{\leftPoint_i -s}{\leftPoint_i-\rightPoint_j}\right) \leftData_i^T \rightDir_j  + \left(\frac{s-\rightPoint_j}{\leftPoint_i-\rightPoint_j}\right) \leftDir_i^T\rightData_j,
\end{displaymath}
so we have immediately, 
$\bfe_i^T \widetilde{\mathcal{K}}(\leftPoint_i) =\leftDir_i^T\RightData=\leftDir_i^T\Cred $ and  $\widetilde{\mathcal{K}}(\rightPoint_j)\bfe_j = \LeftData^T \rightDir_j = \Bred \rightDir_j$. Similarly, for the case $\leftPoint_i=\rightPoint_i$, we obtain
\begin{displaymath}
	\bfe_i^T \widetilde{\mathcal{K}}(\leftPoint_i) =\leftDir_i^T\Cred,\qquad \widetilde{\mathcal{K}}(\rightPoint_i)\bfe_i = \Bred \rightDir_i,
	\qquad\text{and}\qquad
	\bfe_i^T \widetilde{\mathcal{K}}'(\leftPoint_i) \bfe_i = -\mathbb{L}_{i,i} = -\theta_i.
\end{displaymath}

The remainder of this section is structured as follows. In \Cref{subsec:N2} we consider the special case $\numFunctions=2$ and show its close relation to the Loewner framework. If $\numFunctions\geq3$ we need a strategy to fix the remaining degrees of freedom. To this end we propose two approaches which both provide interpolation of further data while maintaining the dimension of the matrices in the realization. The first approach uses additional interpolation points for this (\Cref{subsec:additionalData}), while the second one interpolates additional derivative evaluations of the transfer functions (\Cref{subsec:derivativeInformation}).

\subsection{Structured Loewner Realizations: The case $\numFunctions=2$}
\label{subsec:N2}
Setting $P_{\LeftData} = P_{\RightData} = I_\numData$ gives $2\numData^2 + (m+p)\numData$ equations in \eqref{eq:Pequations} and \eqref{eq:matrixConditions} for the $\numFunctions\numData^2 + (m+p)\numData$ unknowns such that we can expect (under some regularity) a unique solution for the case $\numFunctions=2$. In this case $\Bred = \LeftData^T$, $\Cred = \RightData$, and the matrix equations in \eqref{eq:matrixConditions} reduce to
\begin{displaymath}
	h_1(\LeftPoint)\Ared_1 + h_2(\LeftPoint)\Ared_2 = \LeftDir^T\RightData\qquad\text{and}\qquad
	\Ared_1 h_1(\RightPoint) + \Ared_2 h_2(\RightPoint) = \LeftData^T\RightDir.
\end{displaymath}
To decouple these equations, we multiply the first equation from the right by $h_2(\RightPoint)$ and the second equation from the left by $h_2(\LeftPoint)$. Subtracting the resulting systems yields the Sylvester-like equation
\begin{equation}
	\label{eq:SylvN21}
	h_2(\LeftPoint)\Ared_1 h_1(\RightPoint) - h_1(\LeftPoint)\Ared_1 h_2(\RightPoint) = h_2(\LeftPoint)\LeftData^T\RightDir - \LeftDir^T\RightData h_2(\RightPoint).
\end{equation}
Similarly, we can eliminate $\Ared_1$ and obtain
\begin{equation}
	\label{eq:SylvN22}
	h_1(\LeftPoint)\Ared_2 h_2(\RightPoint) - h_2(\LeftPoint)\Ared_2 h_1(\RightPoint) = h_1(\LeftPoint)\LeftData^T\RightDir - \LeftDir^T\RightData h_1(\RightPoint).
\end{equation}

\begin{remark}
	If the desired model is a generalized state space system as in \eqref{eq:stateSpaceSystem}, i.\,e., $h_1(s) = s$ and $h_2(s)\equiv -1$, then \eqref{eq:SylvN21} and \eqref{eq:SylvN22} are given by the Sylvester equations
	\begin{align}
		\label{SylvesterEquationsLoewner}
		\Ared_1 \RightPoint - \LeftPoint\Ared_1 = \LeftData^T\RightDir - \LeftDir^T\RightData\qquad\text{and}\qquad
		\Ared_2 \RightPoint - \LeftPoint\Ared_2 =  \LeftPoint \LeftData^T\RightDir-\LeftDir^T\RightData\RightPoint,
	\end{align}
	respectively. Up to a sign factor, these are exactly the Sylvester equations that define the Loewner matrix and the shifted Loewner matrix \cite{MayA07}. In particular, if $\rightPoint_i\neq \leftPoint_j$ for $i,j=1,\ldots,\numData$, then $\Ared_1 = -\mathbb{L}$ and $\Ared_2 = -\mathbb{L}_\sigma$ are the unique solutions of \eqref{eq:SylvN21} and \eqref{eq:SylvN22} and the Loewner framework is a special case of the general framework presented in this paper. Similarly, the proportional ansatz for the realization of delay systems introduced in \cite{SchU15} is covered by our framework.
\end{remark}
Those elements of $\Ared_1$ and $\Ared_2$, for which $\leftPoint_i\neq\rightPoint_j$, may be obtained by multiplying \eqref{eq:SylvN21} and \eqref{eq:SylvN22} from  left by $\bfe_i^T$ and from  right by $\bfe_j$ yielding
\begin{align}
	\label{generalizedLoewnerMatrixa}
	\left[\Ared_1\right]_{i,j} = \frac{h_2(\leftPoint_i)\leftData_i^T\rightDir_j - \leftDir_i^T \rightData_j h_2(\rightPoint_j)}{h_2(\leftPoint_i)h_1(\rightPoint_j)-h_1(\leftPoint_i)h_2(\rightPoint_j)},\quad
	\left[\Ared_2\right]_{i,j} = \frac{h_1(\leftPoint_i)\leftData_i^T\rightDir_j - \leftDir_i^T\rightData_j h_1(\rightPoint_j)}{h_1(\leftPoint_i)h_2(\rightPoint_j)-h_2(\leftPoint_i)h_1(\rightPoint_j)}
\end{align}
under the generic assumption that $h_1(\leftPoint_i)h_2(\rightPoint_j) \neq h_2(\leftPoint_i)h_1(\rightPoint_j)$. This is  satisfied for all possible choices of interpolation points (with $\leftPoint_i\neq\rightPoint_j$) if the functions $h_1$ and $h_2$ satisfy the \emph{Haar condition} \cite{Che82}, see also  \Cref{subsec:additionalData}. The components for which $\leftPoint_i = \rightPoint_i$ can be obtained by translating the conditions in \Cref{thm:interpolationConditions} to the $\numFunctions=2$ case. 
This yields
\begin{equation}
\label{eq:sylvesterForSigmaEqualToMu}
\begin{aligned}
h_1(\leftPoint_i) [\Ared_1]_{i,i}+h_2(\leftPoint_i) [\Ared_2]_{i,i} &=\leftDir_i^T\rightData_i,\\
h'_1(\leftPoint_i) [\Ared_1]_{i,i}+h'_2(\leftPoint_i) [\Ared_2]_{i,i} &=-\theta_i
\end{aligned}
\end{equation}
and consequently
\begin{align}
	\label{generalizedLoewnerMatrixb}
	\left[\Ared_1\right]_{i,i} = \frac{h_2(\leftPoint_i)\theta_i+h'_2(\leftPoint_i)\leftDir_i^T\rightData_i}{h'_2(\leftPoint_i)h_1(\leftPoint_i)-h'_1(\leftPoint_i)h_2(\leftPoint_i)},\quad
	\left[\Ared_2\right]_{i,i} = \frac{h_1(\leftPoint_i)\theta_i+h'_1(\leftPoint_i)\leftDir_i^T\rightData_i}{h'_1(\leftPoint_i)h_2(\leftPoint_i)-h'_2(\leftPoint_i)h_1(\leftPoint_i)},
\end{align}
for the components with $\leftPoint_i = \rightPoint_i$ under the generic assumption  $h'_2(\leftPoint_i)h_1(\leftPoint_i)\neq h'_1(\leftPoint_i)h_2(\leftPoint_i)$.
Consequently, we have proven the subsequent result.
\begin{corollary}
	\label{thm:GeneralizedLoewnerN2}
	Let $\Ared_1$ and $\Ared_2$ be as in \eqref{generalizedLoewnerMatrixa} and \eqref{generalizedLoewnerMatrixb} where the  denominators are assumed nonzero. If
	\begin{displaymath}
		\det\left(h_1(\tilde{s})\Ared_1 + h_2(\tilde{s})\Ared_2\right)\neq 0\qquad\text{for all } \tilde{s}\in\{\leftPoint_i\}_{i=1}^n\cup\{\rightPoint_i\}_{i=1}^n,
	\end{displaymath}
	then the transfer function $\Hred(s) = \RightData\left(h_1(s)\Ared_1 + h_2(s)\Ared_2\right)^{-1}\LeftData^T$ satisfies the interpolation conditions \eqref{eq:allInterpolationConditions}. 
\end{corollary}
The matrices $\Ared_1$ and $\Ared_2$ have a structure similar to the Loewner matrix and the shifted Loewner matrix. This gives rise to the idea that the result of \Cref{thm:GeneralizedLoewnerN2} can be obtained from the standard Loewner framework using transformed data.
\begin{corollary}
	\label{cor:TransformedDataLoewner}
	Suppose that $h_2(\RightPoint)$ and $h_2(\LeftPoint)$ are nonsingular and that the denominators in \eqref{generalizedLoewnerMatrixa} and \eqref{generalizedLoewnerMatrixb} are not zero. Construct the Loewner matrix $\mathbb{L}$ and the shifted Loewner matrix $\mathbb{L}_\sigma$ for the transformed data
\begin{align}
	\mbox{\small left interpolation data:}\ & \left\lbrace\left(\frac{h_1\left(\leftPoint_i\right)}{h_2\left(\leftPoint_i\right)},\frac{\leftDir_i}{h_2\left(\leftPoint_i\right)},\leftData_i\right), i=1,\ldots,\numData\right\rbrace,\nonumber\\	
	\mbox{\small right interpolation data:}\ & \left\lbrace\left(\frac{h_1\left(\rightPoint_i\right)}{h_2\left(\rightPoint_i\right)},\frac{\rightDir_i}{h_2\left(\rightPoint_i\right)},\rightData_i\right), i=1,\ldots,\numData\right\rbrace,	 	\label{eq:transformedInterpolationData}	 \\
		\mbox{\small bitangential derivative data:}\ & \left\lbrace\left(i,\frac{h_2\left(\leftPoint_i\right)\theta_i+h'_2\left(\leftPoint_i\right)\leftDir_i^T\rightData_i}{h'_1\left(\leftPoint_i\right)h_2\left(\leftPoint_i\right)-h'_2\left(\leftPoint_i\right)h_1\left(\leftPoint_i\right)}\right)\ 
		\mbox{\small  for}\ \leftPoint_i=\rightPoint_i\right\rbrace.	\nonumber
\end{align}
	If $\det(h_2(\tilde{s})\mathbb{L}_\sigma - h_1(\tilde{s})\mathbb{L})\neq0$ for all $\tilde{s}\in\{\leftPoint_i\}_{i=1}^n\cup\{\rightPoint_i\}_{i=1}^n$, then the transfer function
	\begin{displaymath}
		\Hred(s) = \RightData(h_2(s)\mathbb{L}_\sigma - h_1(s)\mathbb{L})^{-1}\LeftData^T
	\end{displaymath} 
	interpolates the data.
\end{corollary}

\begin{proof}
Simple calculations yield that, when constructing the Loewner pencil with the transformed interpolation data \eqref{eq:transformedInterpolationData}, the Loewner matrix and the shifted Loewner matrix coincide with $-\Ared_1$ and $\Ared_2$ given in \eqref{generalizedLoewnerMatrixa} and \eqref{generalizedLoewnerMatrixb}. Corollary \ref{thm:GeneralizedLoewnerN2} completes the proof.
\end{proof}

\Cref{cor:TransformedDataLoewner} allows one to transfer many results of the standard Loewner framework to the general framework considered in this subsection. In particular, this allows us to keep the system matrices real if the interpolation data is closed under complex conjugation. The details are formulated in \Cref{lem:realRealizationForKEqual2}.

\begin{lemma}
\label{lem:realRealizationForKEqual2}
Let the interpolation data be closed under complex conjugation, i.\,e., there exist unitary matrices $T_{\LeftData},T_\RightData\in\mathbb{C}^{n\times n}$ with
\begin{alignat*}{3}
&T_{\LeftData}^*\LeftPoint T_{\LeftData}\in\mathbb{R}^{n\times n},\quad &&T_{\LeftData}^*\LeftDir^T\in\mathbb{R}^{n},\quad &&T_{\LeftData}^*\LeftData^T\in\mathbb{R}^{n},\\
&T_{\RightData}^*\RightPoint T_{\RightData}\in\mathbb{R}^{n\times n},\quad &&\RightDir T_{\RightData}\in\mathbb{R}^{n},\quad &&\RightData T_{\RightData}\in\mathbb{R}^{n}.
\end{alignat*}
Moreover, assume that the $\theta_i$'s (for the case $\mu_i=\lambda_i$) are closed under complex conjugation. Then, the realization  $(T_{\LeftData}^*\Ared_1T_{\RightData},\,T_{\LeftData}^*\Ared_2T_{\RightData},\,T_{\LeftData}^*\LeftData^T,\,\RightData T_{\RightData})$ with $(\Ared_1,\,\Ared_2,\,\LeftData^T,\,\RightData)$ from \Cref{thm:GeneralizedLoewnerN2} consists of real-valued matrices and interpolates the data.
\end{lemma}

\begin{proof}
First note that if the interpolation data is closed under complex conjugation, so is the transformed data in \Cref{cor:TransformedDataLoewner}. Based on this observation, the proof for the case that $\mu_i\neq\lambda_j$ for all $i,j=1,\ldots,n$ simply follows the lines of \cite[section~2.4.4.]{AntLI15}. This can also be comprehended after multiplying the Sylvester-like equations \eqref{eq:SylvN21} and \eqref{eq:SylvN22} from  left by $T_{\LeftData}^*$ and from  right by $T_{\RightData}$. Similar reasoning proves the claim for the $\mu_i=\lambda_i$ case.
\end{proof}

\begin{example}
\label{ex:realMatrices}
A special case of \Cref{lem:realRealizationForKEqual2} applies when the interpolation data is sorted such that the real values have the highest indices, i.\,e.,
\begin{align*}
\LeftPoint &= \mathrm{diag}(\leftPoint_1,\,\overline{\leftPoint_1},\,\ldots,\,\leftPoint_{2\ell-1},\,\overline{\leftPoint_{2\ell-1}},\,\leftPoint_{2\ell+1},\,\ldots,\,\leftPoint_{\numData}),\\
\LeftDir &= \begin{bmatrix}
\leftDir_1 & \overline{\leftDir_1} & \ldots & \leftDir_{2\ell-1} & \overline{\leftDir_{2\ell-1}} & \leftDir_{2\ell+1} & \ldots & \leftDir_{\numData}
	\end{bmatrix}, \\
		\LeftData &= \begin{bmatrix}
\leftData_1 & \overline{\leftData_1} & \ldots & \leftData_{2\ell-1} & \overline{\leftData_{2\ell-1}} & \leftData_{2\ell+1} & \ldots & \leftData_{\numData}
	\end{bmatrix},\\ 
 \RightPoint &= \mathrm{diag}(\rightPoint_1,\,\overline{\rightPoint_1},\,\ldots,\,\rightPoint_{2r-1},\,\overline{\rightPoint_{2r-1}},\,\rightPoint_{2r+1},\,\ldots,\,\rightPoint_{\numData}), \\
\RightDir &= \begin{bmatrix}
\rightDir_1 & \overline{\rightDir_1} & \ldots & \rightDir_{2r-1} & \overline{\rightDir_{2r-1}} & \rightDir_{2r+1} & \ldots & \rightDir_{\numData}
	\end{bmatrix},\\
	\RightData &= \begin{bmatrix}
\rightData_1 & \overline{\rightData_1} & \ldots & \rightData_{2r-1} & \overline{\rightData_{2r-1}} & \rightData_{2r+1} & \ldots & \rightData_{\numData}
	\end{bmatrix}.
\end{align*}
In this case possible choices for $T_{\LeftData}$ and $T_{\RightData}$ are given by block diagonal matrices
\begin{equation*}
T_{\bullet} = \mathrm{blkdiag}\left(\frac{1}{\sqrt{2}}\begin{bmatrix}
1 & -\imath\\ 1 & \imath
\end{bmatrix},\,\ldots,\,\frac{1}{\sqrt{2}}\begin{bmatrix}
1 & -\imath\\ 1 & \imath
\end{bmatrix},\,1,\,\ldots,\,1\right),
\end{equation*}
where $\bullet\in\{\LeftData,\RightData\}$. One can also obtain the real realization directly from \Cref{thm:interpolationConditions} by choosing $P_\LeftData^T = T_{\LeftData}^*$ and $P_\RightData = T_{\RightData}$ (see discussion after \Cref{thm:interpolationConditions}).
\end{example}


\begin{remark}
	The result from \Cref{cor:TransformedDataLoewner} can (formally) be obtained by rewriting the transfer function, as similar to what is done in \cite{PonPS15}, namely
		\begin{equation*}
		\Hred(s) = \Cred(h_1(s)\Ared_1+h_2(s)\Ared_2)^{-1}\Bred = \Cred\left(\frac{h_1(s)}{h_2(s)}\Ared_1 + \Ared_2\right)^{-1}\Bred\frac{1}{h_2(s)}.
	\end{equation*}
\end{remark}
%

\subsection{Structured Realization for the Case $\numFunctions \geq 3$}
\label{subsec:K3}

When  $\numFunctions \geq 3$, the conditions in \Cref{thm:interpolationConditions} do not provide  enough conditions for the available degrees of freedom (even if $P_\LeftData$ and $P_\RightData$ are fixed). Hence, we have some freedom in choosing the matrices $\Ared_{k}$ with $k=1,\ldots,\numData$. We can exploit these degrees of freedom, for instance, by fitting the transfer function to additional data. For simplicity we assume $\left\lbrace\leftPoint_i\right\rbrace_{i=1}^\numData\cap\left\lbrace\rightPoint_i\right\rbrace_{i=1}^\numData=\emptyset$ for the remainder of this section.

\subsubsection{Interpolation at Additional Points}
\label{subsec:additionalData}

In this subsection we focus on fitting the transfer function to additional data or, equivalently, match the given data with a smaller state space dimension. To this end, let us assume that we have $(Q_{\LeftData}-1)\numData$ additional left interpolation points and $(Q_{\RightData}-1)\numData$ additional right interpolation points at hand, which we group in sets of $\numData$. More precisely, the left interpolation data is grouped into the matrices
\begin{subequations}
\label{eq:partitionedData}
\begin{equation}
	\begin{gathered}
	\LeftPoint_{q} \vcentcolon= \diag(\leftPoint_{q;1},\leftPoint_{q;2},\ldots,\leftPoint_{q;\numData})\in\mathbb{C}^{\numData\times\numData},\qquad 
	\LeftDir_{q} \vcentcolon= \begin{bmatrix}
		\leftDir_{q;1}&\leftDir_{q;2}&\cdots&\leftDir_{q;\numData}
	\end{bmatrix}\in\mathbb{C}^{p\times\numData},\\
	\LeftData_{q} \vcentcolon= \begin{bmatrix}
		\leftData_{q;1}&\leftData_{q;2}&\cdots&\leftData_{q;\numData}
	\end{bmatrix}\in\mathbb{C}^{m\times\numData},
	\end{gathered}
\end{equation}
where $q=1,\ldots,Q_{\LeftData}$. Here, we set $\leftPoint_{1;i}\vcentcolon=\leftPoint_i$, $\leftData_{1;i}\vcentcolon=\leftData_i$, and $\leftDir_{1;i}\vcentcolon=\leftDir_i$, such that we have $\LeftPoint_1 = \LeftPoint$, $\LeftDir_1 = \LeftDir$, and $\LeftData_1 = \LeftData$. Similarly, we introduce, for $q=1,\ldots,Q_{\RightData}$, the matrices
\begin{equation}
	\begin{gathered}
	\RightPoint_{q} \vcentcolon= \diag(\rightPoint_{q;1},\rightPoint_{q;2},\ldots,\rightPoint_{q;\numData})\in\mathbb{C}^{\numData\times\numData},\qquad 
	\RightDir_{q} \vcentcolon= \begin{bmatrix}
		\rightDir_{q;1}&\rightDir_{q;2}&\cdots&\rightDir_{q;\numData}
	\end{bmatrix}\in\mathbb{C}^{m\times\numData},\\
	\RightData_{q} \vcentcolon= \begin{bmatrix}			
		\rightData_{q;1}&\rightData_{q;2}&\cdots&\rightData_{q;\numData}
	\end{bmatrix}\in\mathbb{C}^{p\times\numData}.
	\end{gathered}
\end{equation}
\end{subequations}
To use the full capacity of available degrees of freedom, we assume $\numFunctions = Q_{\LeftData}+Q_{\RightData}$, with $Q_{\LeftData},Q_{\RightData}\geq1$. The next result gives us the necessary and sufficient conditions that the matrices in the realization $\Hred(s)$ must satisfy to interpolate all prescribed information.

\begin{theorem}
	\label{thm:MatchingAdditionalData}
	Let $\Hred(s) = \Cred\Kred(s)^{-1}\Bred$ with $\Kred(s) = \sum_{k=1}^\numFunctions h_k(s) \Ared_k$ 
	and suppose that $\Kred(s)$ is nonsingular for all $\tilde{s}\in\{\leftPoint_{q;i}\}_{q=1}^{Q_{\LeftData}}\cup \{\rightPoint_{q;i}\}_{q=1}^{Q_{\RightData}}$ for all $i=1,\ldots,\numData$.
	\begin{enumerate}
		\item The left interpolation conditions $\leftDir_{q;i}^T \Hred(\leftPoint_{q;i}) = \leftData_{q;i}^T$ are satisfied for $i=1,\ldots,\numData$ and $q=1,\ldots,Q_{\LeftData}$ if and only if there exist matrices $P_{\LeftData,q}$ with $q=1,\ldots,Q_{\LeftData}$ that satisfy
		\begin{equation}
			\label{eq:conditionsLeft}
			\LeftData_{q}^T = P_{\LeftData,q}^T\Bred \qquad\text{and}\qquad \sum_{k=1}^\numFunctions h_k(\LeftPoint_{q})P_{\LeftData,q}^T\Ared_{k} = \LeftDir_{q}^T\Cred.
		\end{equation}		
		\item The right interpolation conditions $\Hred(\rightPoint_{q;i})\rightDir_{q;i} = \rightData_{q;i}$ are satisfied for $i=1,\ldots,\numData$ and $q=1,\ldots,Q_{\RightData}$ if and only if there exist matrices $P_{\RightData,q}$ with $q=1,\ldots,Q_{\RightData}$ that satisfy
		\begin{equation}
			\label{eq:conditionsRight}
			\RightData_{q} = \Cred P_{\RightData,q}\qquad\text{and}\qquad \sum_{k=1}^\numFunctions \Ared_{k}P_{\RightData,q}h_k(\RightPoint_{q}) = \Bred \RightDir_{q}.
		\end{equation}
	\end{enumerate}		
\end{theorem}

\begin{proof}
	The result follows directly from \Cref{thm:interpolationConditions}. For the sake of completeness we give the proof of the first statement again. The second identity in \eqref{eq:conditionsLeft} implies $\leftDir_{q;i}^T\Cred = \bfe_i^TP_{\LeftData,q}^T\sum_{k=1}^\numFunctions h_k(\mu_{q;i}) \Ared_k$. Thus, by the first identity and the definition of $\Hred$ we conclude
	\begin{displaymath}
		\leftDir_{q;i}^T \Hred(\leftPoint_{q;i}) = \bfe_i^TP_{\LeftData,q}^T\Bred = \leftData_{q;i}^T
	\end{displaymath}
	for $i=1,\ldots,\numData$ and $q=1,\ldots,Q_{\LeftData}$.
\end{proof}
Evidently, in order to satisfy the interpolation conditions \eqref{eq:interpolationCondition} it will be sufficient to require that \eqref{eq:conditionsLeft} and \eqref{eq:conditionsRight} hold simultaneously. This gives us the following strategy to determine the realization matrices $\Ared_k, \Bred$, and $\Cred$. Suppose we can find matrices $P_{\LeftData,q}$ and $P_{\RightData,q}$ that satisfy the first identity in \eqref{eq:conditionsLeft} and \eqref{eq:conditionsRight}, respectively, i.\,e., that allow us to fix $\Bred$ and $\Cred$. 
%
Then we can compute the matrices $\Ared_{k}$ as follows. Vectorization of the second identity in \eqref{eq:conditionsLeft} yields
\begin{displaymath}
	\sum_{k=1}^\numFunctions \left(I_\numData\otimes h_k(\LeftPoint_{q})P_{\LeftData,q}^T\right)\text{vec}(\Ared_{k}) = \left(\Cred^T\otimes I_\numData\right)\text{vec}(\LeftDir_{q}^T),
\end{displaymath}
where $\otimes$ denotes the Kronecker product and $\text{vec}(X)$ denotes the vector of stacked columns of the matrix $X$.
Similarly, we obtain from \eqref{eq:conditionsRight} the equation
\begin{displaymath}
	\sum_{k=1}^\numFunctions \left(h_k(\RightPoint_{q}) P_{\RightData,q}^T\otimes I_\numData\right)\text{vec}(\Ared_{k}) = \left(I_\numData\otimes \Bred\right)\text{vec}(\RightDir_q).
\end{displaymath}
All equations together yield the linear algebraic system $\mathbb{A}\bfalpha = \bfbeta$ with $\mathbb{A}\in\mathbb{C}^{\numFunctions \numData^2\times\numFunctions\numData^2}$, $\bfalpha, \bfbeta \in\mathbb{C}^{\numFunctions\numData^2}$ given by
\begin{equation}
\label{eq:linearSystemAdditionalData}
\begin{aligned}
	\mathbb{A} &\vcentcolon= \begin{bmatrix}
		I_\numData\otimes h_1\left(\LeftPoint_{1}\right)P_{\LeftData,1}^T & \cdots & I_\numData\otimes h_\numFunctions\left(\LeftPoint_{1}\right)P_{\LeftData,1}^T\\
		\vdots & & \vdots\\
		I_\numData\otimes h_1\left(\LeftPoint_{Q_{\LeftData}}\right)P_{\LeftData,Q_{\LeftData}}^T & \cdots & I_\numData\otimes h_\numFunctions\left(\LeftPoint_{Q_{\LeftData}}\right)P_{\LeftData,Q_{\LeftData}}^T\\[.5em]\hline\\[-.9em]
		h_1\left(\RightPoint_{1}\right) P_{\RightData,1}^T\otimes I_\numData & \cdots & h_\numFunctions\left(\RightPoint_{1}\right) P_{\RightData,1}^T\otimes I_\numData\\
		\vdots & & \vdots\\
		h_1\left(\RightPoint_{Q_{\RightData}}\right) P_{\RightData,Q_{\RightData}}^T\otimes I_\numData & \cdots & h_\numFunctions\left(\RightPoint_{Q_{\RightData}}\right) P_{\RightData,Q_{\RightData}}^T\otimes I_\numData
	\end{bmatrix},\\
	\bfalpha &\vcentcolon=  \begin{bmatrix}
		\text{vec}(\Ared_1)\\
		\vdots\\
		\text{vec}(\Ared_\numFunctions)
	\end{bmatrix},\qquad\text{and}\qquad
	\bfbeta \vcentcolon= \begin{bmatrix}
		\left(\Cred^T\otimes I_\numData\right)\text{vec}\left(\LeftDir_{1}^T\right)\\
		\vdots\\
		\left(\Cred^T\otimes I_\numData\right)\text{vec}\left(\LeftDir_{Q_{\LeftData}}^T\right)\\[.5em]\hline\\[-.9em]
		\left(I_\numData\otimes \Bred\right)\text{vec}\left(\RightDir_{1}\right)\\
		\vdots\\
		\left(I_\numData\otimes \Bred\right)\text{vec}\left(\RightDir_{Q_{\RightData}}\right)
	\end{bmatrix}.
	\end{aligned}
\end{equation}
Note that the solution of the linear equation system $\mathbb{A}\bfalpha = \bfbeta$ depends on $P_{\LeftData,q}$ and $P_{\RightData,q}$ and there is some freedom in choosing these matrices. A simple possibility is given by
\begin{equation}
\label{eq:choiceOfP_MIMO}
	P_{\LeftData,q}^T \vcentcolon= \begin{bmatrix}
		\LeftData_{q}^T & *
	\end{bmatrix},\qquad P_{\RightData,q} \vcentcolon= \begin{bmatrix}
		\RightData_{q}\\ *
	\end{bmatrix},\qquad
	\Bred \vcentcolon= \begin{bmatrix}
		I_m\\0
	\end{bmatrix},
	\qquad\text{and}\qquad 
	\Cred \vcentcolon= \begin{bmatrix}
		I_p & 0
	\end{bmatrix}
\end{equation}
which satisfies the first identity in \eqref{eq:conditionsLeft} and \eqref{eq:conditionsRight} for any choice of $*$. However, the trivial choice of setting these blocks to zero makes the system matrix $\mathbb{A}$ singular. Instead, we propose to fill the $*$ part of the matrices $P_{\LeftData,q}$ and $P_{\RightData,q}$ such that they are nonsingular assuming that $\LeftData_{q}$ and $\RightData_{q}$ have full row rank. A more specific choice of $*$ may even lead to real-valued realizations as stated in the following lemma.

\begin{lemma}
\label{lem:realRealizationForAdditionalDataMIMO}
Let each of the interpolation data sets be closed under complex conjugation, i.\,e., there exist unitary matrices $T_{\LeftData,q},T_{\RightData,q}\in\mathbb{C}^{n\times n}$ with
\begin{alignat*}{4}
&T_{\LeftData,q}^*\LeftPoint_q T_{\LeftData,q}\in\mathbb{R}^{n\times n},\quad &&T_{\LeftData,q}^*\LeftDir_q^T\in\mathbb{R}^{n},\quad &&T_{\LeftData,q}^*\LeftData_q^T\in\mathbb{R}^{n},\quad &&\mbox{for}\quad q=1,\ldots,Q_{\LeftData},\\
&T_{\RightData,q}^*\RightPoint_q T_{\RightData,q}\in\mathbb{R}^{n\times n},\quad &&\RightDir_q T_{\RightData,q}\in\mathbb{R}^{n},\quad &&\RightData_q T_{\RightData,q}\in\mathbb{R}^{n},\quad &&\mbox{for}\quad q=1,\ldots,Q_{\RightData}.
\end{alignat*}
Moreover, let the matrices $P_{\LeftData,q}$ and $P_{\RightData,q}$ be as in \eqref{eq:choiceOfP_MIMO} with free entries $*$ chosen such that $T_{\LeftData,q}^*P_{\LeftData,q}^T\in\mathbb{R}^{n}$ and $P_{\RightData,q} T_{\RightData,q}\in\mathbb{R}^{n}$ hold. Then, the matrices $\Ared_1$, ..., $\Ared_\numFunctions$, $\Bred$, and $\Cred$ from \Cref{thm:MatchingAdditionalData} are real matrices (if existent).
\end{lemma}

\begin{proof}
From \eqref{eq:choiceOfP_MIMO} it is clear that $\Bred$ and $\Cred$ are real matrices. In addition, the second equalities in \eqref{eq:conditionsLeft} and \eqref{eq:conditionsRight} are equivalent to
\begin{align*}
\sum_{k=1}^\numFunctions T_{\LeftData,q}^*h_k(\LeftPoint_{q})T_{\LeftData,q}T_{\LeftData,q}^*P_{\LeftData,q}^T\Ared_{k} &= T_{\LeftData,q}^*\LeftDir_{q}^T\Cred\\ \mbox{and}\qquad\sum_{k=1}^\numFunctions \Ared_{k}P_{\RightData,q}T_{\RightData,q}T_{\RightData,q}^*h_k(\RightPoint_{q})T_{\RightData,q} &= \Bred \RightDir_{q}T_{\RightData,q}.
\end{align*}
Since the $\Ared_k$ are the solutions of these linear matrix equations and since their coefficient matrices as well as the right hand sides are real-valued, the $\Ared_k$'s are also real-valued.
\end{proof}

To complete the discussion, we analyze the regularity of $\mathbb{A}$ in the SISO case, that is $p=m=1$. Here, we set 
\begin{equation}
\label{eq:choiceOfPSISO}
	P_{\LeftData,q} \vcentcolon= \diag(\LeftData_{q}),\qquad P_{\RightData,q} \vcentcolon= \diag(\RightData_{q}),\qquad \Bred \vcentcolon= \begin{bmatrix}
		1 \\ \vdots \\ 1
	\end{bmatrix},\quad\text{and}\quad\Cred \vcentcolon= \begin{bmatrix}
		1 & \ldots & 1
	\end{bmatrix}.
\end{equation}
With these settings, the $(i,j)$ components of the second matrix equations in \eqref{eq:conditionsLeft} and \eqref{eq:conditionsRight} read as $\leftData_{q;i} \sum_{k=1}^\numFunctions h_k(\leftPoint_{q;i}) [\Ared_k]_{i,j} = 1$ and $\rightData_{q;j} \sum_{k=1}^\numFunctions h_k(\rightPoint_{q;j}) [\Ared_k]_{i,j} = 1$, respectively. Putting this into matrix notation yields the linear system
\begin{equation}
	\label{eq:HaarSystem}
	\begin{bmatrix}
		\leftData_{1;i} & & & & &\\
		& \ddots &  & & &\\
		& & \leftData_{Q_\LeftData;i} & & &\\
		& & & \rightData_{1;j} & & \\
		& & & & \ddots & \\
		& & & & & \rightData_{Q_\RightData;j}
	\end{bmatrix}\begin{bmatrix}
		h_1(\leftPoint_{1;i}) & \ldots & h_\numFunctions(\leftPoint_{1;i})\\
		\vdots & & \vdots\\
		h_1(\leftPoint_{Q_{\LeftData};i}) & \ldots & h_\numFunctions(\leftPoint_{Q_{\LeftData};i}) \\
		h_1(\rightPoint_{1;j}) & \ldots & h_\numFunctions(\rightPoint_{1;j})\\
		\vdots & & \vdots\\
		h_1(\rightPoint_{Q_{\RightData};j}) & \ldots & h_\numFunctions(\rightPoint_{Q_{\RightData};j})
	\end{bmatrix}\begin{bmatrix}
		[\Ared_1]_{i,j}\\
		[\Ared_2]_{i,j}\\
		\vdots\\
		[\Ared_\numFunctions]_{i,j}
	\end{bmatrix} = \begin{bmatrix}
		1\\1\\\vdots\\1
	\end{bmatrix},
\end{equation}
where the system matrix is the product of a diagonal matrix and a generalized Vandermonde matrix. This generalized Vandermonde matrix is also called a \emph{Haar matrix} and is nonsingular if the driving frequencies $\leftPoint_{q;i}$ and $\rightPoint_{q;j}$ are distinct and the functions $h_k$ satisfy the \emph{Haar condition} \cite{Che82}.  In particular, the Haar condition is satisfied for monomials, and thus relevant for second-order systems (cf. \Cref{tab:StructureExamples}). The diagonal matrix is nonsingular if the driving frequencies $\leftPoint_{q;i}$ and $\rightPoint_{q;j}$ are distinct from the roots of the original transfer function. In this case, the system above has a unique solution for each $(i,j)$ combination and hence, via transformations, we can infer that $\mathbb{A}$ is nonsingular.

We illustrate the construction of the realization with additional data with the following toy example.
\begin{example}
\label{ex:toyExample}
	Given scalars $a_1,a_2,a_3,b,c\in\mathbb{R}$ with $bc\neq0$, consider the system
	\begin{align*}
		a_1\dot{x}(t) &= a_2x(t) + a_3x(t-1) + bu(t),\\
		y(t) &= cx(t)
	\end{align*}
	with transfer function $H(s) = \frac{cb}{sa_1 - a_2 - \mathrm{e}^{-s}a_3}$. Setting $Q_\LeftData = 1$ and $Q_\RightData = 2$, we pick distinct interpolation points $\leftPoint_{1;1} = \leftPoint$, $\rightPoint_{1;1} = \rightPoint$, and $\rightPoint_{2;1} = \lambda$. We choose $\Bred = 1$ and $\Cred=1$ with $P_{\LeftData,1} = H(\leftPoint), P_{\RightData,1} = H(\rightPoint)$, and $P_{\RightData,2} = H(\lambda)$. Then the system in \eqref{eq:HaarSystem} reads as
	\begin{equation}
		\label{eq:exToyExampleSys}
		\begin{bmatrix}
			H(\leftPoint) & & \\
			& H(\rightPoint) & \\
			& & H(\lambda)
		\end{bmatrix}\begin{bmatrix}
			\leftPoint & -1 & -\exp(-\leftPoint)\\
			\rightPoint & -1 & -\exp(-\rightPoint)\\
			\lambda & -1 & -\exp(-\lambda)
		\end{bmatrix}\begin{bmatrix}
			\Ared_1\\
			\Ared_2\\
			\Ared_3
		\end{bmatrix} = \begin{bmatrix}
			1\\1\\1
		\end{bmatrix}.
	\end{equation}
	The inverse of the Haar matrix is given by
	\begin{displaymath}
		\resizebox{.95\hsize}{!}{$
		\frac{1}{\leftPoint\e{\leftPoint}(\e{\rightPoint} - \e{\lambda}) + \rightPoint\e{\rightPoint}(\e{\lambda}-\e{\leftPoint}) + \lambda\e{\lambda}(\e{\leftPoint}-\e{\rightPoint})}
		\begin{bmatrix}
			\e{\leftPoint}(\e{\rightPoint}-\e{\lambda}) & -\e{\rightPoint}(\e{\leftPoint}-\e{\lambda}) & \e{\lambda}(\e{\leftPoint}-\e{\rightPoint})\\
			\e{\leftPoint}(\rightPoint\e{\rightPoint} - \lambda\e{\lambda}) & -\e{\rightPoint}(\leftPoint\e{\leftPoint}-\lambda\e{\lambda}) & \e{\lambda}(\leftPoint\e{\leftPoint}-\rightPoint\e{\rightPoint})\\
			-\e{\leftPoint}\e{\rightPoint}\e{\lambda}(\rightPoint-\lambda) & \e{\leftPoint}\e{\rightPoint}\e{\lambda}(\leftPoint-\lambda) & -\e{\leftPoint}\e{\rightPoint}\e{\lambda}(\leftPoint-\rightPoint)
		\end{bmatrix}$}
	\end{displaymath}
	such that the solution of \eqref{eq:exToyExampleSys} is given by $\begin{bmatrix}
		\Ared_1 & \Ared_2 & \Ared_3
	\end{bmatrix} = \frac{1}{cb}\begin{bmatrix}
		a_1 & a_2 & a_3
	\end{bmatrix}$. In particular, we recover the original transfer function.
\end{example}

Clearly, the realization is real-valued if all quantities in \eqref{eq:HaarSystem} are real. If we pick the driving frequencies on the imaginary axis, then in general the Haar matrix will be complex-valued. The following lemma shows how to obtain real-valued realizations based on complex interpolation data with $P$ matrices as in \eqref{eq:choiceOfPSISO}.

\begin{lemma}
\label{lem:realRealizationForAdditionalDataSISO}
Let the interpolation data be closed under complex conjugation and sorted as in Example \ref{ex:realMatrices} such that the unitary matrices $T_{\LeftData},T_{\RightData}\in\mathbb{C}^{n\times n}$ from Example \ref{ex:realMatrices} satisfy
\begin{alignat*}{4}
&T_{\LeftData}^*\LeftPoint_q T_{\LeftData}\in\mathbb{R}^{n\times n},\quad &&T_{\LeftData}^*\LeftDir_q^T\in\mathbb{R}^{n},\quad &&T_{\LeftData}^*\LeftData_q^T\in\mathbb{R}^{n},\quad &&\mbox{for}\quad q=1,\ldots,Q_{\LeftData},\\
&T_{\RightData}^*\RightPoint_q T_{\RightData}\in\mathbb{R}^{n\times n},\quad &&\RightDir_q T_{\RightData}\in\mathbb{R}^{n},\quad &&\RightData_q T_{\RightData}\in\mathbb{R}^{n},\quad &&\mbox{for}\quad q=1,\ldots,Q_{\RightData}.
\end{alignat*}
Moreover, let the matrices $P_{\LeftData,q}$ and $P_{\RightData,q}$ be as in \eqref{eq:choiceOfPSISO}. Then, the realization 
$$(T_{\LeftData}^*\Ared_1T_{\RightData},\,\ldots,\,T_{\LeftData}^*\Ared_\numFunctions T_{\RightData},\,T_{\LeftData}^*\Bred,\,\Cred T_{\RightData}),$$
with $(\Ared_1,\,\ldots,\,\Ared_\numFunctions,\,\Bred,\,\Cred)$ from \Cref{thm:MatchingAdditionalData}, consists of real-valued matrices and interpolates the data.
\end{lemma}

\begin{proof}
First note that the state space transformation by the unitary matrices $T_{\LeftData}^*$ and $T_{\RightData}$ does not change the transfer function and thus the interpolation given by \Cref{thm:MatchingAdditionalData} is also valid here. It remains to show that the realization consists of real-valued matrices. Since $\Bred$ and $\Cred$ are given in \eqref{eq:choiceOfPSISO}, it is straightforward to see that $T_{\LeftData}^*\Bred$ and $\Cred T_{\RightData}$ are real-valued.
As in the proof of \Cref{lem:realRealizationForAdditionalDataMIMO}, we deduce the realness of  $T_{\LeftData}^*\Ared_k T_{\RightData}$ by observing  that the second equalities in \eqref{eq:conditionsLeft} and \eqref{eq:conditionsRight} are equivalent to
\begin{align*}
\sum_{k=1}^\numFunctions T_{\LeftData}^*h_k(\LeftPoint_{q})T_{\LeftData}T_{\LeftData}^*P_{\LeftData,q}^TT_{\LeftData}T_{\LeftData}^*\Ared_{k}T_{\RightData} &= T_{\LeftData}^*\LeftDir_{q}^T\Cred T_{\RightData}\\ \mbox{and}\qquad\sum_{k=1}^\numFunctions T_{\LeftData}^*\Ared_{k}T_{\RightData}T_{\RightData}^*P_{\RightData,q}T_{\RightData}T_{\RightData}^*h_k(\RightPoint_{q})T_{\RightData} &= T_{\LeftData}^*\Bred \RightDir_{q}T_{\RightData}.
\end{align*}
Straightforward computations yield that $T_{\LeftData}^*P_{\LeftData,q}^TT_{\LeftData}$ and $T_{\RightData}^*P_{\RightData,q}T_{\RightData}$ are real-valued. From these linear matrix equations we can determine the $\Ared_k$ or equivalently their transformed analogues $T_{\LeftData}^*\Ared_kT_{\RightData}$. In the latter case, we observe that the coefficient matrices as well as the right hand sides are real-valued and thus the $T_{\LeftData}^*\Ared_kT_{\RightData}$ are also real-valued.
\end{proof}


A crucial point in \Cref{thm:MatchingAdditionalData} is the nonsingularity of the affine structure $\Kred(s) = \sum_{k=1}^\numFunctions h_k(s) \Ared_k$ at the driving frequencies $\leftPoint_{q;i}$ and $\rightPoint_{q;i}$. However, if we add more and more data we expect that at some point the information become redundant, and hence $\Kred(s)$ might become singular. To remove the redundant part, we suppose that 
\begin{equation}
	\label{eq:redundantAss}
	\mathrm{rank} \left(\sum_{k=1}^\numFunctions h_k(s)\Ared_k\right) = \mathrm{rank}\left(\begin{bmatrix}
		\Ared_1 & \cdots & \Ared_\numFunctions
	\end{bmatrix}\right) = \mathrm{rank}\left(\begin{bmatrix}
		\Ared_1 \\ \vdots \\ \Ared_\numFunctions
	\end{bmatrix}\right) =\vcentcolon r
\end{equation}
holds for all $s\in\{\leftPoint_{q;i}\}\cup\{\rightPoint_{q;i}\}$. In this case there exist unitary matrices $V = \begin{bmatrix}
	V_1 & V_2
\end{bmatrix}$ and $W = \begin{bmatrix}
	W_1 & W_2
\end{bmatrix}\in\mathbb{C}^{\numData\times\numData}$ with $V_1,W_1\in\mathbb{C}^{\numData\times r}$ and $V_2,W_2\in\mathbb{C}^{\numData\times (\numData-r)}$ such that 
\begin{equation}
	\label{eq:nullSpaceMatrices}
		\Ared_kV_2 = 0 \qquad\text{and}\qquad
		\Ared_k^*W_2 = 0,\qquad\text{for all}\quad k=1,\ldots,\numFunctions.
\end{equation}

\begin{theorem}
	\label{thm:truncationAdditionalData}
	Let the realization $\Hred(s) = \Cred(\sum_{k=1}^\numFunctions h_k(s)\Ared_k)^{-1}\Bred$ satisfy the equations in \Cref{thm:MatchingAdditionalData} with matrices $P_{\LeftData,q}$ and $P_{\RightData,q}$. Suppose that the $\Ared_k$'s satisfy the rank assumption \eqref{eq:redundantAss} and let $V_1,W_1\in\mathbb{C}^{\numData\times r}$ complete $V_2$ and $W_2$ in \eqref{eq:nullSpaceMatrices} to unitary matrices. For $k=1,\ldots,\numFunctions$ set
	\begin{displaymath}
		\Ared_{k;r} \vcentcolon= W_1^*\Ared_k V_1,\qquad \Bred_r \vcentcolon= W_1^*\Bred, \qquad\text{and}\qquad \Cred_r \vcentcolon = \Cred V_1.
	\end{displaymath}	
If $\spann \{\leftDir_{q;1},\,\ldots,\, \leftDir_{q;\numData}\}=\mathbb{C}^p$ for all $q=1,\ldots,Q_\LeftData$ and $\spann \{\rightDir_{q;1},\,\ldots,\, \rightDir_{q;\numData}\}=\mathbb{C}^m$ for all $q=1,\ldots,Q_\RightData$, then the realization $\Hred_r(s) = \Cred_r(\sum_{k=1}^\numFunctions h_k\left(s\right)\Ared_{k;r})^{-1}\Bred_r$ interpolates the data.	
\end{theorem}

\begin{proof}
	First, bear in mind that by the assumption, the affine structure $\sum_{k=1}^\numFunctions h_k(s)\Ared_{k;r}$ is nonsingular at the driving frequencies $\leftPoint_{q;i}$ and $\rightPoint_{q;i}$, and observe that $\Ared_k V_1 V_1^* = \Ared_k$ and $W_1W_1^*\Ared_k = \Ared_k$ hold for $k=1,\ldots,\numFunctions$ by construction of $V_1$ and $W_1$. Thus, for $q=1,\ldots,Q_\LeftData$
	\begin{align*}
		\sum_{k=1}^\numFunctions h_k(\LeftPoint_q)P_{\LeftData,q}^TW_1 \Ared_{k;r} &= \left(\sum_{k=1}^\numFunctions h_k(\LeftPoint_q)P_{\LeftData,q}^T W_1W_1^*\Ared_k\right)V_1\\
		 &= \left(\sum_{k=1}^{\numFunctions} h_k(\LeftPoint_q) P_{\LeftData,q}^T\Ared_k\right)V_1 = \LeftDir_q^T\Cred_r,
	\end{align*}
	where the second identity follows from \eqref{eq:conditionsLeft}. Similarly, we obtain for $q=1,\ldots,Q_\RightData$
	\begin{align*}
		\sum_{k=1}^\numFunctions \Ared_{k;r} V_1^* P_{\RightData,q}h_k(\RightPoint_q) &=  W_1^*\sum_{k=1}^\numFunctions\Ared_k V_1V_1^* P_{\RightData,q} h_k(\RightPoint_q) = W_1^* \sum_{k=1}^\numFunctions \Ared_k P_{\RightData,q}h_k(\RightPoint_q) = \Bred_r \RightDir_q.
	\end{align*}
	Furthermore, notice 
	\begin{align*}
		\LeftDir_q^T\Cred = \sum_{k=1}^\numFunctions h_k(\LeftPoint_q)P_{\LeftData,q}^T\Ared_k = \sum_{k=1}^\numFunctions h_k(\LeftPoint_q)P_{\LeftData,q}^T\Ared_k V_1V_1^* = \LeftDir_q^T\Cred V_1V_1^*.
	\end{align*}
	Since the columns of $\LeftDir_q$ span the whole space $\mathbb{C}^p$, the above identity implies $\Cred = \Cred V_1V_1^*$. With the same reasoning we obtain $\Bred = W_1W_1^* \Bred$. Finally, we have
	\begin{align*}
		\leftDir_{q;i}^T \Hred_r(\leftPoint_{q;i}) &= \bfe_i^T \LeftDir_q^T\Cred_r\left(\sum_{k=1}^\numFunctions h_k(\leftPoint_{q;i})\Ared_{k;r}\right)^{-1}\Bred_r\\
		&= \bfe_i^T \left(\sum_{k=1}^\numFunctions h_k(\LeftPoint_q)P_{\LeftData,q}^TW_1\Ared_{k;r}\right)\left(\sum_{k=1}^{\numFunctions} h_k(\leftPoint_{q;i})\Ared_{k;r}\right)^{-1}\Bred_r\\
		&= \bfe_i^T P_{\LeftData,q}^TW_1\left(\sum_{k=1}^\numFunctions h_k(\leftPoint_{q;i})\Ared_{k;r}\right)\left(\sum_{k=1}^\numFunctions h_k(\leftPoint_{q;i})\Ared_{k;r}\right)^{-1}\Bred_r \\
		&= \bfe_i^T P_{\LeftData,q}^T W_1W_1^* \Bred = \leftData_{q;i}^T
	\end{align*}
	for $q=1,\ldots,Q_\LeftData$ and $i=1,\ldots,\numData$. The right interpolation conditions follow analogously.
\end{proof}

\begin{example}
	If we pick further distinct interpolation points in \Cref{ex:toyExample}, then the realization is given by the matrices
	\begin{displaymath}
		\Ared_1 = \frac{1}{cb}\begin{bmatrix}
			a_1 & \ldots & a_1\\
			\vdots &  & \vdots\\
			a_1 & \ldots & a_1
		\end{bmatrix}, \quad \Ared_2 = \frac{1}{cb}\begin{bmatrix}
			a_2 & \ldots & a_2\\
			\vdots &  & \vdots\\
			a_2 & \ldots & a_2
		\end{bmatrix}, \quad\text{and}\quad \Ared_3 = \frac{1}{cb}\begin{bmatrix}
			a_3 & \ldots & a_3\\
			\vdots &  & \vdots\\
			a_3 & \ldots & a_3
		\end{bmatrix}.
	\end{displaymath}
	Clearly, the rank assumption \eqref{eq:redundantAss} is satisfied with $r=1$. Setting $W_1 = \begin{bmatrix}
		1 & 0 & \ldots & 0]
	\end{bmatrix}$ and $V_1 = W_1^T$ yields the true transfer function.
\end{example}

\subsubsection{Matching Derivative Data}
\label{subsec:derivativeInformation}
Hermite interpolation provides a well known and robust approach for polynomial approximation that involves the matching of derivative data.    
When we seek reduced models that are structurally equivalent to standard first order realizations (that is, when we have 
 in \eqref{eq:FOMstructured} $\numFunctions = 2$, $h_1(s) = s$, and $h_2(s) \equiv -1$)
 then first order necessary conditions for optimality of the reduced order approximant with respect to the $\mathcal{H}_2$ norm are known and they require that the reduced transfer function $\Hred(s)$ must be a Hermite interpolant of the original $H(s)$\cite{GuGAB08}. 
Even though these necessary conditions do not extend immediately to more general structured systems as appear in \eqref{eq:FOMstructured}, it is known for some special cases
such as second order systems with modal damping and port-Hamiltonian systems \cite{BeaB14}, and for systems with simple delay structures \cite{PonPS15,PonGBPS16}, that Hermite interpolation (in a different form then for the rational case) still plays a fundamental role in the necessary optimality conditions. Therefore, if derivative information for the transfer function $H(s)$ is accessible then this motivates finding a structurally equivalent realization $\Hred(s)$ that matches both the evaluation data 
and the derivative data. 
 Assume that we have
\begin{equation}
	\label{eq:hermiteCondition}
	\leftDir_i^T H'(\leftPoint_i) = (\leftData_i')^T\qquad\text{and}\qquad H'(\rightPoint_i)\rightDir_i = \rightData_i'\qquad\text{for } i=1,\ldots,\numData
\end{equation}
available, where $H'$ denotes the derivative of $H$, i.\,e., $H' := \frac{\mathrm{d}}{\mathrm{ds}}H$, and $(\leftData_i')^T$ and $\rightData_i'$ are the tangential interpolation values of $H'$. These are collected in the matrices
\begin{displaymath}
	\LeftData' = \begin{bmatrix}
		\leftData_1' & \ldots & \leftData_\numData'
	\end{bmatrix}\qquad\text{and}\qquad
	\RightData' = \begin{bmatrix}
		\rightData_1' & \ldots & \rightData_\numData'
	\end{bmatrix}.
\end{displaymath}
In this section, we derive conditions such that the transfer function $\Hred$ interpolates the data \eqref{eq:interpolationData} with $\left\lbrace\leftPoint_i\right\rbrace_{i=1}^\numData\cap\left\lbrace\rightPoint_i\right\rbrace_{i=1}^\numData=\emptyset$ and satisfies in addition the Hermite interpolation condition \eqref{eq:hermiteCondition}. 

\begin{theorem}
	\label{thm:hermiteInterpolation}
	Let $\Hred(s) = \Cred(\sum_{k=1}^{\numFunctions} h_k(s)\Ared_k)^{-1}\Bred$ and suppose that $\sum_{k=1}^\numFunctions h_k(\tilde{s})\Ared_k$ is nonsingular for all $\tilde{s}\in\left\lbrace\leftPoint_i\right\rbrace_{i=1}^\numData\cup\left\lbrace\rightPoint_i\right\rbrace_{i=1}^\numData$.
	\begin{enumerate}
		\item The left interpolation conditions $\leftDir_i^T \Hred(\leftPoint_i) = \leftData_i^T$ and the left Hermite interpolation conditions $\leftDir_i^T \Hred'(\leftPoint_i) = \left(\leftData_i'\right)^T$ are satisfied for $i=1,\ldots,\numData$ if and only if there exist matrices $P_{\LeftData}$ and $P_{\LeftData'}$ that satisfy
			\begin{align}
				\label{eq:leftHermite1}\LeftData^T &= P_{\LeftData}^T \Bred, & \sum_{k=1}^\numFunctions h_k(\LeftPoint)P_{\LeftData}^T \Ared_{k} &= \LeftDir^T \Cred,\\
				\label{eq:leftHermite2}\left(\LeftData'\right)^T &= \left(P_{\LeftData'}\right)^T \Bred, & \sum_{k=1}^\numFunctions h_k(\LeftPoint)\left(P_{\LeftData'}\right)^T \Ared_{k} &= -\sum_{k=1}^\numFunctions h_k'(\LeftPoint)P_{\LeftData}^T \Ared_{k}.
			\end{align}
		\item The right interpolation conditions $\Hred(\rightPoint_i)\rightDir_i = \rightData_i$ and the right Hermite interpolation conditions $\Hred'(\rightPoint_i)\rightDir_i = \rightData_i'$ for $i=1,\ldots,\numData$ are satisfied if and only if there exist matrices $P_{\RightData}$ and $P_{\RightData'}$ that satisfy
			\begin{align}
				\label{eq:rightHermite1}\RightData &= \Cred P_{\RightData}, & \sum_{k=1}^\numFunctions \Ared_{k} P_{\RightData}h_k(\RightPoint) &= \Bred \RightDir,\\
				\label{eq:rightHermite2}\RightData' &= \Cred P_{\RightData'}, & \sum_{k=1}^\numFunctions \Ared_{k}P_{\RightData'} h_k(\RightPoint) &= -\sum_{k=1}^\numFunctions \Ared_{k} P_{\RightData} h_k'(\RightPoint).
		\end{align}
	\end{enumerate}
\end{theorem}

\begin{proof}
	We only prove the first statement;  the second statement is proved analogously. Observe that \eqref{eq:leftHermite1} are the left interpolation conditions from \Cref{thm:interpolationConditions}, $\leftDir_i^T H(\leftPoint_i) = \leftData_i^T$ for $i=1,\ldots,\numData$. It remains to show that the left Hermite interpolation conditions are equivalent to \eqref{eq:leftHermite2}. As before, let  $\widetilde{\mathcal{K}}(s)=\sum_{k=1}^\numFunctions h_k(s)\Ared_{k}$. The second identity in \eqref{eq:leftHermite1} holds if and only if
	\begin{align*}
		\leftDir_i^T \Cred = e_i^T \LeftDir^T \Cred = e_i^T P_{\LeftData}^T \widetilde{\mathcal{K}}(\leftPoint_i)\qquad\mbox{for}\;i=1,\ldots,n.
	\end{align*}
Similarly from the second identity in \eqref{eq:leftHermite2} we obtain
	\begin{align*}
		-e_i^T P_{\LeftData}^T\widetilde{\mathcal{K}}'(\leftPoint_i) &= -e_i^T \sum_{k=1}^\numFunctions h_k'(\LeftPoint)P_{\LeftData}^T\Ared_{k} = e_i^T \sum_{k=1}^\numFunctions h_k(\LeftPoint)\left(P_{\LeftData'}\right)^T\Ared_{k} = e_i^T \left(P_{\LeftData'}\right)^T \widetilde{\mathcal{K}}(\leftPoint_i).
	\end{align*}
	Thus, for $i=1,\ldots,\numData$ we have
	\begin{align*}
		\leftDir_i^T \Hred'(\leftPoint_i) &= -\leftDir_i^T \Cred \widetilde{\mathcal{K}}(\leftPoint_i)^{-1}\widetilde{\mathcal{K}}'(\leftPoint_i)\widetilde{\mathcal{K}}(\leftPoint_i)^{-1}\Bred = -e_i^TP_{\LeftData}^T\widetilde{\mathcal{K}}'(\leftPoint_i)\widetilde{\mathcal{K}}(\leftPoint_i)^{-1}\Bred\\
		&= e_i^T \left(P_{\LeftData'}\right)^T \Bred = \left(\leftData_i'\right)^T,
	\end{align*}
	where the last identity is nothing else than the first equality in \eqref{eq:leftHermite2}.
\end{proof}
As before, it is sufficient and necessary to satisfy \eqref{eq:leftHermite1}-\eqref{eq:rightHermite2} simultaneously to satisfy the interpolation conditions \eqref{eq:interpolationCondition} and the Hermite interpolation conditions \eqref{eq:hermiteCondition}. After fixing the matrices $P_\LeftData$, $P_{\LeftData'}$, $P_{\RightData}$, and $P_{\RightData'}$, \Cref{thm:hermiteInterpolation} gives $4\numData^2$ equations for $\numFunctions\numData^2$ unknown variables. In particular for $\numFunctions=4$, we can expect under some regularity conditions that there is a unique solution for the matrices $\Ared_{k}$. Hereby, the $P$ matrices can be chosen similarly as in the previous section, for example as
\begin{equation}
	\label{eq:choiceOfP_MIMOHermite}
	P_{\LeftData}^T \vcentcolon= \begin{bmatrix}
		\LeftData^T & *
	\end{bmatrix},\qquad 
	P_{\LeftData'}^T \vcentcolon= \begin{bmatrix}
		\left(\LeftData'\right)^T & *
	\end{bmatrix},\qquad 
	P_{\RightData} \vcentcolon= \begin{bmatrix}
		\RightData\\ *
	\end{bmatrix},\qquad
	P_{\RightData'} \vcentcolon= \begin{bmatrix}
		\RightData'\\ *
	\end{bmatrix},
\end{equation} 
yielding $\Bred = \begin{bmatrix}
	I_m & 0
\end{bmatrix}^T$ and $\Cred = \begin{bmatrix}
	I_p & 0
\end{bmatrix}$.
For $\numFunctions=3$, we can either satisfy the left or the right Hermite interpolation conditions. 
For the sake of completeness, we derive the equivalent of the system \eqref{eq:linearSystemAdditionalData} for Hermite interpolation for $\numFunctions=4$. Vectorization of the second equations in \eqref{eq:leftHermite1}-\eqref{eq:rightHermite2}, respectively, yields the system $\mathbb{A}\bfalpha = \bfbeta$ with matrix $\mathbb{A}\in\mathbb{C}^{4\numData^2,4\numData^2}$ and vectors $\bfalpha,\bfbeta\in\mathbb{C}^{4\numData^2}$ given by
\begin{equation}
\label{eq:linearSystemHermiteData}
\begin{aligned}
	\mathbb{A}&\vcentcolon=\begin{bmatrix}
		I_n\otimes h_1(\LeftPoint)P_{\LeftData}^T & \ldots & I_n\otimes h_4(\LeftPoint)P_{\LeftData}^T\\
		I_n\otimes \left(h_1(\LeftPoint)\left(P_{\LeftData'}\right)^T + h_1'(\LeftPoint)P_{\LeftData}^T\right) & \ldots & I_n\otimes \left(h_4(\LeftPoint)\left(P_{\LeftData'}\right)^T + h_4'(\LeftPoint)P_{\LeftData}^T\right)\\
		h_1(\RightPoint)P_{\RightData}^T \otimes I_n & \ldots & h_4(\RightPoint)P_{\RightData}^T \otimes I_n\\
		\left(h_1(\RightPoint)\left(P_{\RightData'}\right)^T + h_1'(\RightPoint)P_{\RightData}^T\right)\otimes I_n & \ldots & \left(h_4(\RightPoint)\left(P_{\RightData'}\right)^T + h_4'(\RightPoint)P_{\RightData}^T\right)\otimes I_n
	\end{bmatrix},\\
	\bfalpha &\vcentcolon= \begin{bmatrix}
		\mathrm{vec}(\Ared_{1})\\ 
		\mathrm{vec}(\Ared_{2})\\
		\mathrm{vec}(\Ared_{3})\\
		\mathrm{vec}(\Ared_{4})
	\end{bmatrix},\qquad\text{and}\qquad 
	\bfbeta \vcentcolon= \begin{bmatrix}
		\left(\Cred^T\otimes I_n\right)\mathrm{vec}\left(\LeftDir^T\right)\\
		0\\
		\left(I_n\otimes \Bred\right)\mathrm{vec}\left(\RightDir\right)\\
		0
	\end{bmatrix}.
	\end{aligned}
\end{equation}
\begin{remark}
	\label{rem:realRealizationForHermite}
Real-valued realizations that accomplish Hermite interpolation may be obtained in the same manner as in the case of additional interpolation points (cf. Lemmas \ref{lem:realRealizationForAdditionalDataMIMO} and \ref{lem:realRealizationForAdditionalDataSISO}). The only additional requirement is that $\LeftData$ and $\LeftData'$ as well as $\RightData$ and $\RightData'$ need to have the same number of complex conjugate pairs such that
\begin{equation*}
T_{\LeftData}^*\LeftData^T\in\mathbb{R}^{n},\quad T_{\LeftData}^*\left(\LeftData'\right)^T\in\mathbb{R}^{n},\quad \RightData T_{\RightData}\in\mathbb{R}^n,\quad\mbox{and}\quad \RightData' T_{\RightData}\in\mathbb{R}^n.
\end{equation*}
\end{remark}

Suppose we have solved the linear system \eqref{eq:linearSystemHermiteData} to obtain the realization $\Hred(s) = \Cred\Kred(s)^{-1}\Bred$ with $\Kred(s) = \sum_{k=1}^\numFunctions h_k(s)\Ared_k$. By construction, the matrices satisfy the equations in \Cref{thm:hermiteInterpolation}. However, $\Kred(s)$ might be singular at the driving frequencies $\leftPoint_i$ and $\rightPoint_i$. If the rank condition \eqref{eq:redundantAss} is satisfied, then we can truncate the redundant data as in \Cref{thm:truncationAdditionalData}, i.\,e., we construct matrices $\Ared_{k;r}, \Bred_r$, and $\Cred_r$ yielding the reduced realization $\Hred_r(s) = \Cred_r(\sum_{k=1}^\numFunctions h_k(s)\Ared_{k;r})^{-1}\Bred_r$. This model still satisfies the left and right interpolation conditions by \Cref{thm:truncationAdditionalData} and it suffices to check if the Hermite interpolation conditions are matched as well. By the same reasoning as in the proof of \Cref{thm:truncationAdditionalData} we can establish the identity
\begin{displaymath}
	\sum_{k=1}^\numFunctions h_k(\LeftPoint)\left(P_{\LeftData'}\right)^T W_1 \Ared_{k;r} = -\sum_{k=1}^\numFunctions h_k'(\LeftPoint)P_{\LeftData}^TW_1\Ared_{k;r}
\end{displaymath}
and compute
\begin{align*}
	\leftDir_i^T \Hred_r'(\leftPoint_i) &= -\bfe_i^T \LeftDir^T \Cred_r \left(\sum_{k=1}^\numFunctions h_k(\leftPoint_i)\Ared_{k;r}\right)^{-1} \left(\sum_{k=1}^\numFunctions h_k'(\leftPoint_i)\Ared_{k;r}\right) \left(\sum_{k=1}^\numFunctions h_k(\leftPoint_i)\Ared_{k;r}\right)^{-1}\Bred_r\\
	&= -\bfe_i^T P_{\LeftData}^T W_1 \left(\sum_{k=1}^{\numFunctions} h_k'(\leftPoint_i)\Ared_{k;r}\right)\left(\sum_{k=1}^\numFunctions h_k(\leftPoint_i)\Ared_{k;r}\right)^{-1}\Bred_r\\
	&= \bfe_i^T \left(\sum_{k=1}^\numFunctions h_k(\LeftPoint)\left(P_{\LeftData'}\right)^T W_1 \Ared_{k;r}\right)\left(\sum_{k=1}^\numFunctions h_k(\leftPoint_i)\Ared_{k;r}\right)^{-1}\Bred_r\\
	&= \bfe_i^T \left(P_{\LeftData'}\right)^T W_1 W_1^*\Bred = \bfe_i^T \left(P_{\LeftData'}\right)^T \Bred = \left(\leftData_i'\right)^T
\end{align*}
and hence the left Hermite interpolation condition is still satisfied. The proof for the right Hermite interpolation condition proceeds analogously. We summarize the previous discussion in the following theorem.

\begin{theorem}
	\label{thm:truncationDerivativeData}
	Let the realization $\Hred(s) = \Cred(\sum_{k=1}^\numFunctions h_k(s)\Ared_k)^{-1}\Bred$ satisfy the equations in \Cref{thm:hermiteInterpolation} with matrices $P_{\LeftData}$, $P_{\LeftData'}$, $P_{\RightData}$, and $P_{\RightData'}$. Suppose that the $\Ared_k$'s satisfy the rank assumption \eqref{eq:redundantAss} and let $W_1,V_1\in\mathbb{C}^{\numData\times r}$ be as in \Cref{thm:truncationAdditionalData}. If $\spann \{\leftDir_{1},\,\ldots,\, \leftDir_{\numData}\}=\mathbb{C}^p$ and $\spann \{\rightDir_{1},\,\ldots,\, \rightDir_{\numData}\}=\mathbb{C}^m$, then the realization $\Hred_r(s) = \Cred_r(\sum_{k=1}^\numFunctions h_k\Ared_{k;r})^{-1}\Bred_r$ interpolates the data and derivative data with
	\begin{displaymath}
		\Ared_{k;r} \vcentcolon= W_1^*\Ared_k V_1,\qquad \Bred_r \vcentcolon= W_1^*\Bred, \qquad\text{and}\qquad \Cred_r \vcentcolon = \Cred V_1.
	\end{displaymath}
\end{theorem}
\begin{remark}
		Evidently, as the number of functions $\numFunctions$ determining the structure increases, the number of 
	available degrees of freedom to force interpolation increases as well, and in particular, when $\numFunctions>4$ there will be sufficient degrees of freedom available to allow matching of higher order derivatives as well.  The calculations involved are both annoyingly technical and unenlightening, so we choose not to pursue this thread here. In any case for the applications we have in mind,  $\numFunctions\leq 4$, and Hermite interpolation is seen to provide a satisfactory level of fidelity in the reduced models. 
\end{remark}

\subsection{An Algorithm for Structured Realization}
\label{subsec:realizationAlgorithm}
In this section, we synthesize the results of the previous subsections into an algorithmic format, starting with interpolation data \eqref{eq:interpolationData} and an affine structure given via continuously differentiable functions $h_k$ for $k=1,\ldots,\numFunctions$. The goal is to construct matrices $\Ared_1,\ldots,\Ared_\numFunctions$, $\Bred$ and $\Cred$ such that the realization $\Hred(s) = \Cred(\sum_{k=1}^\numFunctions h_k(s)\Ared_k)^{-1}\Bred$ associated with the affine structure interpolates the data. 
We construct realizations as described in previous designated subsections, taking advantage of the simplifications 
available when $\numFunctions=2$. 
Before doing so, a pre-processing step is included if the data is closed under complex conjugation, which facilitates
 construction of a real-valued realization where appropriate. 
Although in principle the transformation to a real-valued realization could be performed after assembling the matrices, 
it is advisable to enforce this in advance, since rounding errors tend to break the underlying conjugate symmetry 
and will cause drift away from a real-valued realization. 
A post-processing step may also be necessary to truncate redundancies
discovered in the interpolation data. 
Details are summarized in \Cref{alg:structuredRealization}.

\begin{algorithm}[ht]
	\caption{Structured Realization}
	\label{alg:structuredRealization}
	\begin{algorithmic}[1]
		\Statex \textbf{Input:} Interpolation data \eqref{eq:interpolationData}, affine structure $h_1(s),\ldots,h_\numFunctions(s)$ with $\numFunctions\in\mathbb{N}$.
		\Statex \textbf{Output:} Matrices $\Ared_1,\ldots,\Ared_\numFunctions$, $\Bred$, and $\Cred$ such that $\Hred(s) = \Cred(\sum_{k=1}^\numFunctions h(s)\Ared_k)^{-1}\Bred$ interpolates the data
		\Statex
		\If{Data is closed under complex conjugation} \Comment{Keep realization real}
		\State	Transform data as in \Cref{lem:realRealizationForKEqual2}, \Cref{lem:realRealizationForAdditionalDataMIMO}, \Cref{lem:realRealizationForAdditionalDataSISO}, or \Cref{rem:realRealizationForHermite}
		\EndIf
		\Statex
		\If{$\numFunctions = 2$} 
		\State Transform data as in \eqref{eq:transformedInterpolationData}
		\State Construct Loewner matrices according to \eqref{eq:LoewnerMatrix} and \eqref{eq:LoewnerMatrixDeriv} from the transformed data
		\State Set $\Ared_1 = -\mathbb{L}, \Ared_2 = \mathbb{L}_\sigma, \Bred = \LeftData^T$ and $\Cred=\RightData$
		\Else
			\If{derivative data \eqref{eq:hermiteCondition} is available}
				\State Construct $\Bred,\Cred$ and $P$ matrices, for example as in \eqref{eq:choiceOfP_MIMOHermite}
				\State Assemble system \eqref{eq:linearSystemHermiteData} and solve for $\Ared_1,\ldots,\Ared_\numFunctions$
			\Else
				\State Partition the data as in \eqref{eq:partitionedData} and adjust $\numData$ accordingly
				\State Construct $\Bred,\Cred$ and $P$ matrices, for example as in \eqref{eq:choiceOfP_MIMO}
				\State Assemble system \eqref{eq:linearSystemAdditionalData} and solve for $\Ared_1,\ldots,\Ared_\numFunctions$
			\EndIf		
		\EndIf
		\Statex
		\State Compute $r$ as in \eqref{eq:redundantAss}
		\If{$r<\numData$} \Comment{Truncation of redundant data}
			\State Compute $V_1$ and $W_1$ as in \Cref{thm:truncationAdditionalData}
			\State Set $\Ared_k \vcentcolon= W_1^*\Ared_kV_1$, $\Bred \vcentcolon= W_1^*\Bred$, and $\Cred \vcentcolon= \Cred V_1$
		\EndIf
	\end{algorithmic}
\end{algorithm}

\subsection{Connection to Structure-preserving Interpolatory Projections}

Although our focus here is on data-driven interpolation, we revisit briefly the structure-preserving interpolatory projection framework introduced in \cite{BeaG09} and establish a connection with realizations arising from \Cref{thm:GeneralizedLoewnerN2}.

\begin{theorem}[Structure-preserving interpolatory projection \cite{BeaG09}]  
	\label{thm:ProjInt}
Consider the generalized realization $H(s)=\mathcal{C}(s)\mathcal{K}(s)^{-1}\mathcal{B}(s)$ where both $\mathcal{C}(s)\in \mathbb{C}^{p \times\dimFOM}$  and $\mathcal{B}(s)\in \mathbb{C}^{\dimFOM\times m}$ are analytic in the right half plane and $\mathcal{K}(s)\in \mathbb{C}^{\dimFOM \times \dimFOM}$ is analytic and full rank 
throughout the right half plane. Suppose that the left interpolation points $\{\leftPoint_1,\ldots,\leftPoint_\numData\}$ together with the left tangential directions $\{\leftDir_1,\ldots,\leftDir_\numData\}$ and the right interpolation points $\{\rightPoint_1,\ldots,\rightPoint_\numData\}$ together with the right tangential directions $\{\rightDir_1,\ldots,\rightDir_\numData\}$ are given. Define $V\in \mathbb{C}^{\dimFOM\times \numData}$ and  $W \in  \mathbb{C}^{\dimFOM\times \numData}$ as
	\begin{subequations}
	\label{eq:projectionMatrices}
	\begin{equation}
		\label{eq:leftProjection}
		W = [\mathcal{K}(\leftPoint_1)^{-T}\mathcal{C}(\leftPoint_1)^T\leftDir_1, \cdots , \mathcal{K}(\leftPoint_\numData)^{-T}\mathcal{C}(\leftPoint_\numData)^T\leftDir_\numData].
	\end{equation}
	and
	\begin{equation}
		\label{eq:rightProjection}
		V=[\mathcal{K}(\rightPoint_1)^{-1}\mathcal{B}(\rightPoint_1)\rightDir_1, \cdots , \mathcal{K}(\rightPoint_n)^{-1}\mathcal{B}(\rightPoint_n)\rightDir_\numData]
	\end{equation}
	\end{subequations}
	Define
	\begin{equation} 
		\label{proj}
		\begin{aligned} 
			\widetilde{\mathcal{K}}(s)=W^T {\mathcal{K}}(s)V,~~
			\widetilde{\mathcal{B}}(s)=W^T {\mathcal{B}}(s),~~\mbox{and}~~
			\widetilde{\mathcal{C}}(s)={\mathcal{C}}(s)V.
		\end{aligned}
	\end{equation}
Then the reduced transfer function $\Hred(s)=\widetilde{\mathcal{C}}(s)\widetilde{\mathcal{K}}(s)^{-1}\widetilde{\mathcal{B}}(s)$  satisfies the interpolation conditions \eqref{eq:allInterpolationConditions}.
\end{theorem}
If we use $\mathcal{K}\left(s\right)=\sum
	_{k=1}^\numFunctions h_k\left(s\right)A_k$, $\mathcal{B}\left(s\right)=B$, and $\mathcal{C}\left(s\right)=C$ for the affine structure we employ here, in  \Cref{thm:ProjInt}, then \eqref{proj} leads to a reduced model with
		$$
	\widetilde{A}_k = W^T A_k V,~\mbox{for}~k=1,\ldots,K,~\widetilde{B} = W^T B,~~\mbox{and}~~\widetilde{C} = C V.
	$$
The question we want to answer next is how (and if) this projection-based reduced model is connected to the data-driven one we develop here.  The next result provides the link.
\begin{prop}
\label{prop:sylvesterForProjection}
	The projection matrices $W$ and $V$ introduced in \eqref{eq:projectionMatrices}, based on the matrix functions $\mathcal{K}\left(s\right)=\sum_{k=1}^\numFunctions h_k\left(s\right)A_k$, $\mathcal{B}\left(s\right)=B$, and $\mathcal{C}\left(s\right)=C$, satisfy the matrix equations
	\begin{equation}
		\label{eq:MatrixEquationProjection}
		\sum_{k=1}^\numFunctions h_k(\LeftPoint)W^T A_k = \LeftDir^T C\qquad\text{and}\qquad 
		\sum_{k=1}^\numFunctions A_k V h_k(\RightPoint) = B\RightDir,
	\end{equation}
	as well as 
	\begin{equation}
		\sum_{k=1}^\numFunctions h'_k(\leftPoint_i) [W^TA_kV]_{i,i} =-\theta_i
	\end{equation}
for those $i$ with $\leftPoint_i=\rightPoint_i$.
\end{prop}
\begin{proof}
	Let $\bfw_i = W\bfe_i$ and $\bfv_i=V\bfe_i$ denote the columns of the projection matrices $W$ and $V$. For $i=1,\ldots,\numData$ we have
	\begin{align*}
		\bfe_i^T \sum_{k=1}^\numFunctions h_k(\LeftPoint)W^T A_k &= \bfw_i^T\sum_{k=1}^\numFunctions  h_k(\leftPoint_i)A_k = \leftDir_i^T C,
	\end{align*}
	which proves the first identity. The second identity is proven similarly whereas the third identity follows from the definitions of $W$ and $V$ and from
	\begin{displaymath}
		\sum_{k=1}^\numFunctions h'_k(\leftPoint_i) [W^TA_kV]_{i,i} = \bfw_i^T\left(\sum_{k=1}^\numFunctions h'_k(\leftPoint_i)A_k\right)\bfv_i = -\leftDir_i^TH'(\leftPoint_i)\rightDir_i = -\theta_i.
	\end{displaymath}
\end{proof}
 \Cref{prop:sylvesterForProjection} gives a better understanding of the realization of \Cref{thm:GeneralizedLoewnerN2} connecting it to the projection-based MOR framework.
 To make this connection more precise,  we will investigate the $\numFunctions=2$ and 
$\numFunctions \geq 3$ cases separately below.
\subsubsection{The $\numFunctions=2$ case}
Using the identities $W^TB = \LeftData^T$ and $CV = \RightData$, we can rewrite \eqref{eq:SylvN21}, using $\numFunctions=2$, as
\begin{displaymath}
	h_2(\LeftPoint)\Ared_{1} h_1(\RightPoint) - h_1(\LeftPoint)\Ared_{1} h_2(\RightPoint) = h_2(\LeftPoint)W^T B\RightDir - \LeftDir^TCV h_2(\RightPoint).
\end{displaymath}
Substituting the expressions for $B\RightDir$ and $\LeftDir^T C$ from \eqref{eq:MatrixEquationProjection} into the right-hand side implies
\begin{displaymath}
	h_2(\LeftPoint)\Ared_1 h_1(\RightPoint) - h_1(\LeftPoint)\Ared_1 h_2(\RightPoint) = h_2(\LeftPoint)W^T A_1 V h_1(\RightPoint) - h_1(\LeftPoint)W^T A_1 V h_2(\RightPoint),
\end{displaymath}
which establishes the relation $\Ared_1 = W^T A_1 V$ as long as the interpolation sets $\left\lbrace\leftPoint_i\right\rbrace_{i=1}^\numData$ and $\left\lbrace\rightPoint_i\right\rbrace_{i=1}^\numData$ are disjoint. The identity $\Ared_2 = W^T A_2 V$ is obtained by using \eqref{eq:SylvN22} instead of \eqref{eq:SylvN21}. Thus for $\numFunctions=2$, our structured realization approach gives exactly the reduced model one would obtain via projection if the original system matrices were to be available. This equivalence of the projected matrices and the matrices obtained by the realizations is also true if there are overlappings between the left and right interpolation point sets. This may be comprehended by the observation that the projected matrices also satisfy \eqref{eq:sylvesterForSigmaEqualToMu} which is clear due to Proposition \ref{prop:sylvesterForProjection}.

\subsubsection{The $\numFunctions\geq3$ case}
Consider the second-order model $H(s) = C(s^2 A_1 + s A_2 + A_3)^{-1}B$. For simplicity, assume that $H(s)$ is SISO.
Given the $2n$ interpolation points $\{\leftPoint_1,\ldots,\leftPoint_\numData\}$  and  $\{\rightPoint_1,\ldots,\rightPoint_\numData\}$, one can obtain a projection-based reduced model $\Hred(s) = \Cred(s^2 \Ared_1 + s \Ared_2 + \Ared_3)^{-1}\Bred$ using \Cref{thm:ProjInt}. This reduced model will interpolate $H(s)$ at $2n$ interpolation points. However, $\Hred(s)$ has $3n$ degrees of freedom\footnote{A second-order model with $\Hred(s) = (s\Cred_1+\Cred_2)(s^2 \Ared_1 + s \Ared_2 + \Ared_3)^{-1}\Bred$, i.\,e., not only the state $\bfx(t)$ but also the velocity $\dot{\bfx}(t)$ is measured, has $4n$ degrees of freedom. But here we do not consider this case.}
and should be able to satisfy $3n$ interpolation conditions. The projection framework cannot achieve this goal. However, our structured realization framework with either additional data as in
Section \ref{subsec:additionalData} or Hermite interpolation as in Section \ref{subsec:derivativeInformation} will construct a reduced model that can match this maximum number of interpolation conditions. In other words,  for  $\numFunctions\geq3$, the structured realization cannot be obtained via projection and indeed satisfies more interpolation conditions than the projection framework. Next we give a numerical example illustrating this discussion on a delay example.

\begin{example}
	We consider the system with affine structure $h_1(s) = s, h_2(s) \equiv -1$, and $h_3(s) = -\exp(-s)$ and matrices
	\begin{displaymath}
		A_1 = \begin{bmatrix}
			1 & 0\\0 & 2
		\end{bmatrix}, \qquad A_2 = A_3 = \begin{bmatrix}
			1 & 0\\0 & 1
		\end{bmatrix}, \qquad b = \begin{bmatrix}
			1\\1
\end{bmatrix},\qquad\text{and}\qquad c = \begin{bmatrix}
	1 \\ 1
\end{bmatrix}	 
	\end{displaymath}
	with transfer function $H(s) = c^T(sA_1 - A_2 - \mathrm{e}^{-s}A_3)^{-1}b$. We set $Q_\LeftData = 1$ and $Q_\RightData = 2$ and pick the driving frequencies $\leftPoint_{1;1} = 0, \rightPoint_{1;1} = 1$, and $\rightPoint_{2;1}=-1$. We want to make use of the system \eqref{eq:HaarSystem}, i.\,e., we set $\Bred = 1, \Cred = 1, P_{\LeftData,1} = \leftData_{1;1}, P_{\RightData,1} = \rightData_{1;1}$, and $P_{\RightData,2} = \rightData_{2;1}$. Altogether, the solution of the system \eqref{eq:HaarSystem} is given by
	\def\e{\mathrm{e}}
	\begin{displaymath}
		\begin{bmatrix}
			\Ared_1\\
			\Ared_2\\
			\Ared_3
		\end{bmatrix} = \frac{1}{2-\e-\frac{1}{\e}}\begin{bmatrix}
			\e-\frac{1}{\e} + \frac{(1-\frac{1}{\e})^2}{2-\e}+\frac{(\e-1)(\e+2)(\e+3)}{-5-2\e}\\
			-\e-\frac{1}{\e} - \frac{1-\frac{1}{\e}}{\e-2} - \frac{\left(\e+2\right)\left(\e+3\right)}{-5-2\e}\\
			2 + \frac{1-\frac{1}{\e}}{\e-2} + \frac{\left(\e+2\right)\left(\e+3\right)}{-5-2\e}
		\end{bmatrix}.
	\end{displaymath}
	Clearly, $\Ared_2\neq\Ared_3$; and hence the realization cannot be obtained 
	via projection.
\end{example}

\section{Examples}
\label{sec:examples}

To illustrate the consequences of the preceding theoretical discussion, we compare various structured realizations 
against the standard Loewner realization framework, using in each case response data that is presented as in \eqref{eq:interpolationData}.  In all the following examples, $H(s)$, $\Hred_\mathrm{L}(s)$, $\Hred_\mathrm{A}(s)$, and $\Hred_\mathrm{H}(s)$ will denote, respectively: the transfer function of the original model, the rational approximation via the standard Loewner realization, the structured realization interpolating at additional points (\cref{subsec:additionalData}), and the structured realization satisfying Hermite interpolation  conditions(\cref{subsec:derivativeInformation}). In the following plots, we represent interpolation data with circles.  Additional driving frequencies used for the structured realization interpolating additional points are presented as diamonds.


The results presented in the previous sections are valid for the general MIMO case, however, for simplicity, we restrict ourselves to SISO examples. Accordingly, the $P$ matrices needed for the realizations corresponding to $\Hred_\mathrm{A}$ and $\Hred_\mathrm{H}$ have been chosen as in \eqref{eq:choiceOfPSISO} and as the analogue for the Hermite case which is
\begin{equation*}
	P_{\LeftData} \vcentcolon= \diag(\LeftData),\quad P_{\RightData} \vcentcolon= \diag(\RightData),\quad P_{\LeftData'} \vcentcolon= \diag(\LeftData'),\quad \mbox{and}\quad P_{\RightData'} \vcentcolon= \diag(\RightData').
\end{equation*}

\begin{example}
	\label{delayExample}
	We test our approaches with the delay model from \cite{BeaG09} given by the $\dimFOM\times\dimFOM$ matrices
	\begin{displaymath}
		A_1 = \nu I_N + T,\qquad A_2 = \frac{1}{\tau}\left(\frac{1}{\zeta}+1\right)(T-\nu I_N),\qquad A_3 = \frac{1}{\tau}\left(\frac{1}{\zeta}-1\right)(T-\nu I_N),
	\end{displaymath}	
	where $T$ is an $\dimFOM\times\dimFOM$ matrix with ones on the sub- and superdiagonal, at the $(1,1)$, and at the $(\dimFOM,\dimFOM)$ position and zeros everywhere else. The ${h_k}$'s are given by $h_1(s) = s, h_2(s) \equiv -1$, and $h_3(s) = -\mathrm{e}^{-\tau s}$. We choose $\dimFOM = 500$, $\tau=1$, $\zeta=0.01$, and $\nu = 5$. The input matrix $B\in\mathbb{R}^{\dimFOM}$ has ones in the first two components and zeros everywhere else and we choose $C = B^T$. 
	We pick $\numData=4$ logarithmically equidistant points on the imaginary axis between  $1\imath$ and $100\imath$ (indicated as circles in \Cref{fig:delayExampleBodePlot}) together with their complex conjugates. For the additional point framework (\cref{subsec:additionalData}) we set $Q_\LeftData = 1$ and $Q_\RightData=2$, such that we have two additional interpolation points (diamonds in \Cref{fig:delayExampleBodePlot}) plus their complex conjugates. The Bode plots of the transfer functions and of the errors are illustrated for the different approaches in \Cref{fig:delayExampleBodePlot} and \Cref{fig:delayExampleErrorPlot}, respectively.
	\begin{figure}[h]
		\centering
		\begin{subfigure}[b]{\linewidth}
			\centering
			\input{ExampleDelay-bodePlot-4}
			\caption{Bode plot of $H$, $\Hred_{L}$, $\Hred_A$, and $\Hred_H$.}
			\label{fig:delayExampleBodePlot}
		\end{subfigure}\\[1em]
		\begin{subfigure}[b]{\linewidth}
			\centering
			\input{ExampleDelay-errorPlot-4}
			\caption{Bode plot of the absolute errors of $\Hred_{L}$, $\Hred_A$, and $\Hred_H$.}
			\label{fig:delayExampleErrorPlot}
		\end{subfigure}
		\caption{\Cref{delayExample} - Transfer functions of the different realizations with $\numData=4$}
		\label{fig:delayExample}
	\end{figure}
	Both of our approaches capture the dynamics of the full model (the graphs are almost on top of that of the original model) and clearly outperform the Loewner realization. This is supported by the $\mathcal{H}_\infty$ errors for the different realizations presented in \Cref{tab:delayExample} - given also for other choices of $\numData$. 
	\begin{table}[H]
		\centering
		\caption{\Cref{delayExample} - $\mathcal{H}_\infty$ errors of the different realizations}
		\label{tab:delayExample}
		\begin{tabular}{rcccc}
			$\numData$ & Loewner & Additional points & Hermite\\\toprule
			$4$ & 2.342312e-01 & 4.496194e-02 & 4.011660e-02\\
			$6$ & 2.449003e-01 & 5.100268e-02 & 4.116856e-02\\
			$8$ & 3.397454e-01 & 4.673353e-02 & 4.307346e-02\\
			$10$ & 5.561860e-01 & 4.454640e-02 & 3.694951e-02\\\bottomrule
		\end{tabular}
	\end{table}	
	Clearly, the choice of the complex driving frequencies $\leftPoint_i$ and $\rightPoint_j$ is important and should be investigated further, but this is not within the scope of this paper.
\end{example}

\begin{example}[{\Cref{acousticWaveExample} continued}]
	\label{ex:acousticWave}
	We generate data for this model using a model for acoustic transmission in a duct presented by Curtain and Morris in \cite{CurM09}. Based on a PDE model, Curtain and Morris derive an analytic transfer function for this problem: $H(s) = \rho_0 \sinh((L-\xi_0)s/c)/\cosh(Ls/c)$, where $\rho_0$ is the air density.  For our case, we assign parameter values: $L=1$, $\xi_0=1/2$, $c=1$, and $\rho_0=1$ and generate data by sampling the Curtain-Morris transfer function on the imaginary axis between $0.1\imath$ and $10\imath$ (see \Cref{fig:acousticWaveExampleChris}). To keep the realization real we add the complex conjugate driving frequencies.
	 We seek structurally equivalent realizations to the hypothesized structure from \Cref{acousticWaveExample} that will interpolate this data. The frequency response of the original transfer function $H(s)$ together with the different structurally equivalent  realizations is presented in \Cref{fig:acousticWaveExampleChris}. The relative error plot \Cref{fig:acousticWaveExampleChrisRel} shows that structured realizations in this case outperform the Loewner realization by several orders of magnitude.
	\begin{figure}[h]
		\centering
		\begin{subfigure}[b]{\linewidth}
			\centering
			\input{ExampleAcousticWave-bodePlot-16}
			\caption{Bode plot of the acoustic transmission model and the structured realizations}
			\label{fig:acousticWaveExampleChris}
		\end{subfigure}
		\begin{subfigure}[b]{\linewidth}
			\centering
			\input{ExampleAcousticWave-relErrorPlot-16}
			\caption{Relative error plot for the different realizations}
			\label{fig:acousticWaveExampleChrisRel}
		\end{subfigure}
		\caption{\Cref{ex:acousticWave} - Bode and relative error plot for $\numData=16$}
	\end{figure}
	It is noteworthy that the exact transfer function can be written in accordance with the hypothesized structure using matrices $\bfc^T=[\,0\ 0\ 0\ \rho_0\,]$, $\bfb^T=[\,1\ 0\ 0\ 0\,]$ and
	\begin{displaymath}
		A_1=\begin{bmatrix}  
			1 & 0 & 0 & 0\\  
			0 & 1 & 0 & 0\\  
			0 & 0 & 1 & 0\\  
			0 & 1 & 0 & 1
		\end{bmatrix}, \quad 
		A_2=\begin{bmatrix}  
			0 & 0 & 0 & 0\\  
			0 & 0 & -1 & 0\\  
			0 & 0 &  0 & 0\\  
			-1 & 0 & 0 &  0 
		\end{bmatrix}, \quad 
		A_3=\begin{bmatrix}
			0 & 0 & 0 & 0\\
			0 & 0 & 0 & 0\\
			0 & 0 &  0 & -1\\
			-1 & 0 & 0 &  0 
		\end{bmatrix}.
	\end{displaymath}
\end{example}

\begin{example}
	\label{delayExample2}
	A heated rod with distributed control and homogeneous Dirichlet boundary conditions, which is cooled by delayed feedback, can be modeled (cf. \cite{BreMV09,MicJM11}) as
	\begin{equation}
		\label{eq:heatedRod}
		\begin{aligned}
		\frac{\partial v(\xi,t)}{\partial t} &= \frac{\partial^2 v(\xi,t)}{\partial \xi^2} + a_1(\xi)v(\xi,t) + a_2(\xi)v(\xi,t-1) + u(t)\quad &&\text{in } (0,\pi)\times(0,T],\\
		v(0,t) &= v(\pi,t) = 0 &&\text{in } [0,T]
		\end{aligned}
	\end{equation}
	For the coefficient functions we choose $a_1(\xi) = -2\sin(\xi)$ and $a_2(\xi) = 2\sin(\xi)$. Discretization of \eqref{eq:heatedRod} via centered finite differences with step size $h \vcentcolon= \frac{\pi}{N+1}$ yields the system
	\begin{align*}
		\dot{\bfx}(t) &= (L_N+A_{1,N})\bfx(t) + A_{2,N}\bfx(t-1) + B\bfu(t),\\
		\bfy(t) &= C\bfx(t),
	\end{align*}
	where $L_N\in\mathbb{R}^{\dimFOM\times\dimFOM}$ is the discrete Laplacian and $A_{1,\dimFOM},A_{2,\dimFOM}\in\mathbb{R}^{\dimFOM\times\dimFOM}$ are discrete approximations of the functions $a_1$ and $a_2$, respectively. The input matrix $B\in\mathbb{R}^{\dimFOM}$ is a vector of ones. As output we use the average temperature of the rod, i.\,e, $C = \frac{1}{\|B\|}B^T$. %
	\begin{figure}[h]%
		\centering
		\begin{subfigure}[b]{\linewidth}
			\centering
			\input{ExampleDelay2-bodePlot-4}
			\caption{Bode plot of $H$, $\Hred_{L}$, $\Hred_A$, and $\Hred_H$.}
			\label{fig:delayExample2BodePlot}
		\end{subfigure}\\[1em]
		\begin{subfigure}[b]{\linewidth}
			\centering
			\input{ExampleDelay2-errorPlot-4}
			\caption{Bode plot of the absolute errors of $\Hred_{L}$, $\Hred_A$, and $\Hred_H$.}
			\label{fig:delayExample2ErrorPlot}
		\end{subfigure}
		\caption{\Cref{delayExample2} - Transfer functions of the different realizations with $\numData=4$}
		\label{fig:delayExample2}
	\end{figure}%
For our tests we use $\dimFOM=100$ and $\numData=4$ interpolation points on the imaginary axis between $10^{-1}\imath$ and $10^3\imath$ together with their complex conjugates. For the realization obtained by interpolating additional data we use the same settings as in \Cref{delayExample}. 
	\begin{table}[h]
		\centering
		\caption{\Cref{delayExample2} - $\mathcal{H}_\infty$ errors for the different realizations}
		\label{tab:delayExample2}
		\begin{tabular}{rcccc}
			$\numData$ & Loewner & Additional points & Hermite\\\toprule
			$4$ & 5.863023e-01 & 1.596379e-01 & 1.751535e-01\\
			$6$ & 7.118732e-01 & 4.716281e-01 & 7.580182e-02\\
			$8$ & 2.735014e-01 & 3.020142e-02 & 3.725486e-02\\
			$10$ & 2.110771e-01 & 1.796065e-01 & 4.085510e-02\\\bottomrule
		\end{tabular}
	\end{table}
	Similarly as in \Cref{delayExample}, our approaches are the only ones that capture the qualitative behavior of the original system (cf. \Cref{fig:delayExample2}). This is true for all tested numbers of interpolation data $\numData$ and is further illustrated by the $\mathcal{H}_\infty$ errors listed in \Cref{tab:delayExample2}.  For this example the difference is not as striking as in the two preceding examples, which are much harder to approximate 
	with a rational transfer function of low degree.
\end{example}

\begin{example}
	\label{secondOrderExample}	
The full model comes from a finite element discretization of a cantilevered Euler-Bernoulli beam \cite[\S 1.16]{hughes2012finite}, resulting in a second order system having the form 
	\begin{displaymath}
		A_1\ddot{\bfx}(t) + A_2\dot{\bfx}(t) + A_3\bfx(t) = B\bfu(t),\qquad \bfy(t) = C\bfx(t).
	\end{displaymath} 
This is a SISO system ($m=1$ and $p=1$) with $N=400$ internal degrees of freedom. The input $\bfu(t)$ represents                                                               
a point force applied to the state $\bfx_1$ ($B = \bfe_1$), while the output 
is the displacement history at $\bfx_N$ ($C = \bfe_{N}^T$). 
The damping matrix $A_2$ models light proportional damping: $A_2 = \alpha_1 A_1 + \alpha_2 A_2$ with
$\alpha_1 = \alpha_2 = 0.05$.
		The realizations are obtained for $\numData=30$ complex driving frequencies on the imaginary axis between $10^{-5}\imath$ and $10^{2}\imath$ together with their complex conjugates (see upper part of \Cref{fig:secondOrderExample}).	
		 Since the transfer function of the original model is a rational transfer function unlike in the previous example, we expect the Loewner realization to perform close to our proposed approach here, which is indeed the case as illustrated in \Cref{fig:secondOrderExample}. The figure shows that both the Loewner realization and the structured realization with additional interpolation points capture the transfer function of the original model quite accurately.  
	\begin{figure}[h]%
		\centering
		\input{ExampleSecondOrder-bodePlotDistribution-30}
		\caption{\Cref{secondOrderExample} - Bode plot of $\Hred_{L}$ and $\Hred_A$ with $\numData=30$}
		\label{fig:secondOrderExample}
	\end{figure}
	\begin{figure}[H]%
		\centering
		\begin{subfigure}[b]{\linewidth}
			\centering
			\input{ExampleSecondOrder-errorPlot-30}
			\caption{Bode plot of the absolute error of $\Hred_{L}(s)$ and $\Hred_A(s)$}
			\label{fig:secondOrderExampleError}
		\end{subfigure}\\[1em]
		\begin{subfigure}[b]{\linewidth}
			\centering
			\input{ExampleSecondOrder-stabsep-30}
			\caption{Bode plot of the stable and unstable part of $\Hred_{L}(s)$ and $\Hred_A(s)$}
			\label{fig:secondOrderExampleStabsep}
		\end{subfigure}
		\caption{\Cref{secondOrderExample} - Transfer functions of the realizations with $\numData=30$.}
	\end{figure}
	However, the error plot in \Cref{fig:secondOrderExampleError} clearly shows the superior behavior of our approach, especially for higher frequencies; the maximum error due to $\Hred_A(s)$ is one order of magnitude smaller than  the error due to $\Hred_{L}(s)$.
	We conclude with a remark on the stability of the reduced models.  As one expects, stability of the reduced model in the Loewner framework depends on the quality of the interpolation (sampling) points.  The Loewner framework does not guarantee a stable reduced model in general.  For a better selection of points (in some cases, optimal) one can, for example, combine the Loewner framework with interpolatory $\mathcal{H}_2$ optimal methods as done in \cite{BeaG12}.  For cases where the Loewner model is unstable, \cite{GosA16} offers various effective post-processing techniques allowing to extract a stable model while not losing much accuracy. One solution is simply to discard the unstable part of the resulting model. Indeed, this choice can be shown to be the best solution in minimizing an $\mathcal{H}_2$-related distance; see, for example, \cite{MagBG10,Koh14,GosA16} for details. For this beam example, both the Loewner and our approach yield unstable reduced models. Following \cite{GosA16}, we then checked how much the stable and unstable parts of the reduced models contribute to the approximation. For both models, the unstable part has only a minor, negligible contribution as illustrated in \Cref{fig:secondOrderExampleStabsep}, where the frequency response plots for the stable and unstable part of the Loewner realization and the structured realization obtained with additional data are displayed. For this example, simply truncating the anti-stable part of the reduced models and taking only the stable part as the approximation causes almost no loss in accuracy.  
Indeed, for $\Hred_{A}(s)$, while the $\mathcal{L}_\infty$ norm of the antistable part is  $8.8882 \times 10^{-6}$, the $\mathcal{H}_\infty$ norm of the stable part is  $1.8799$.  It appears that this unstable part is due to near non-minimality of the reduced  models. We computed poles and zeros of $\Hred_{A}(s)$, and observed that the unstable poles are  very nearly matched by corresponding zeros as listed in \Cref{tab:secondOrderExample} below:
	\begin{table}[H]
		\centering
		\caption{\Cref{secondOrderExample} - Near pole-zero cancellation }
		\label{tab:secondOrderExample}
		\begin{tabular}{lcl}
			\qquad Poles &~ & \qquad Zeros\\\toprule
		     $ 2.87550 \times 10^{-1}$  &~ & $2.87548 \times 10^{-1}$   \\
      $1.94273 \times 10^{-1}  \pm \imath 2.83074\times 10^{-1}$ &~ &
       $1.94274 \times 10^{-1}  \pm \imath 2.83070\times 10^{-1}$ \\
      $5.39041 \times 10^{-2}  \pm \imath 3.51432 \times 10^{-1}$ & ~ &
        $5.39018 \times 10^{-2}  \pm \imath 3.51431 \times 10^{-1}$ \\
			\bottomrule
		\end{tabular}
	\end{table}
Unlike the case for the Loewner framework, we cannot simply take the stable-part of $\Hred_{A}(s)$ as the approximant, since this truncation is performed  after conversion to first-order form and destroys the structure we are seeking to retain. For many examples, including the previous ones considered here, no equivalent, generic, finite-dimensional, first-order structure exists. Therefore, one might consider modifying  \Cref{alg:structuredRealization} so that these near pole-zero cancellations can be detected during the construction and removed without destroying structure.  This is not the focus of this paper and is deferred to a later work.  
\end{example}

\section{Conclusion}
\label{sec:conclusion}
We have introduced a new framework for structured realizations that are derived from input/output data obtained by measurements of an (unknown) transfer function. The models obtained have the form $\Cred(\sum_{k=1}^\numFunctions h_k(s)\Ared_k)^{-1}\Bred$, which allows for a variety of different structures such as internal delays or second order systems. If the chosen structure is a generalized state space representation then our framework coincides with the Loewner realization \cite{MayA07}. In this sense, our work can be seen as an extension of the Loewner framework to more general system structures. Indeed, for $\numFunctions=2$ we showed that structured realizations can be obtained directly via the Loewner framework with transformed data. Based on necessary and sufficient conditions for interpolation, we have offered two strategies for the more general case $\numFunctions>2$, the first allowing for interpolation at additional interpolation points and the second allowing for additional interpolation of derivative information of the transfer function. The remarkable effectiveness of such structured realizations is demonstrated through several examples. 


\bibliographystyle{plain}
\bibliography{StructuredLoewner}

\end{document}